\documentclass[12pt]{article}

\usepackage[dvipdfmx]{graphicx}
\usepackage{amsfonts}
\usepackage[mathscr]{eucal}
\usepackage{ascmac}
\usepackage{dcolumn}
\usepackage{bm}
\usepackage[colorlinks=true,linkcolor=blue,citecolor=blue]{hyperref}
\usepackage{color}
\usepackage{amsmath}

\begin{document}


\newcommand{\gtrsim}{ \mathop{}_{\textstyle \sim}^{\textstyle >} }
\newcommand{\lesssim}{ \mathop{}_{\textstyle \sim}^{\textstyle <} }
\newcommand{\vev}[1]{ \left\langle {#1} \right\rangle }
\newcommand{\bra}[1]{ \langle {#1} | }
\newcommand{\ket}[1]{ | {#1} \rangle }
\newcommand{\EV}{ \ {\rm eV} }
\newcommand{\KEV}{ \ {\rm keV} }
\newcommand{\MEV}{\  {\rm MeV} }
\newcommand{\GEV}{\  {\rm GeV} }
\newcommand{\TEV}{\  {\rm TeV} }
\newcommand{\1}{\mbox{1}\hspace{-0.25em}\mbox{l}}
\newcommand{\Red}[1]{{\color{red} {#1}}}

\newcommand{\lmk}{\left(}  
\newcommand{\rmk}{\right)}
\newcommand{\lkk}{\left[}  
\newcommand{\rkk}{\right]}
\newcommand{\lhk}{\left \{ }  
\newcommand{\rhk}{\right \} }
\newcommand{\del}{\partial}  
\newcommand{\la}{\left\langle} 
\newcommand{\ra}{\right\rangle}
\newcommand{\half}{\frac{1}{2}}

\newcommand{\bea}{\begin{array}}
\newcommand{\eea}{\end{array}}
\newcommand{\beq}{\begin{eqnarray}}
\newcommand{\eeq}{\end{eqnarray}}

\newcommand{\dd}{\mathrm{d}}
\newcommand{\Mpl}{M_{\rm Pl}}
\newcommand{\mg}{m_{3/2}}
\newcommand{\abs}[1]{\left\vert {#1} \right\vert}
\newcommand{\mphi}{m_{\phi}}
\newcommand{\Hz}{\ {\rm Hz}}
\newcommand{\for}{\quad \text{for }}
\newcommand{\Min}{\text{Min}}
\newcommand{\Max}{\text{Max}}
\newcommand{\Kahler}{K\"{a}hler }
\newcommand{\cphi}{\varphi}

\begin{titlepage}

\baselineskip 8mm

\begin{flushright}
IPMU 15-0069
\end{flushright}

\begin{center}

\vskip 1.2cm

{\Large\bf
Charged Q-balls in gauge mediated SUSY breaking models
}

\vskip 1.8cm

{\large 
Jeong-Pyong Hong$^{a}$, 
Masahiro Kawasaki$^{a,b}$, 
and
Masaki Yamada$^{a,b}$}

\vskip 0.4cm

{\it$^a$Institute for Cosmic Ray Research, The University of Tokyo,
5-1-5 Kashiwanoha, Kashiwa, Chiba 277-8582, Japan}\\
{\it$^b$Kavli IPMU (WPI), UTIAS, The University of Tokyo, 5-1-5 Kashiwanoha, 
Kashiwa, 277-8583, Japan}

\date{\today}
\vspace{2cm}

\begin{abstract}  
It is known that after Affleck-Dine baryogenesis, spatial inhomogeneities of Affleck-Dine field grow into non-topological solitons called Q-balls. In gauge mediated SUSY breaking models, sufficiently large Q-balls with baryon charge are stable while Q-balls with lepton charge can always decay into leptons. For a Q-ball that carries nonzero $B$ and $L$ charges, the difference between the baryonic component and the leptonic component in decay rate may induce nonzero electric charge on the Q-ball. This implies that charged Q-ball, also called gauged Q-ball, may emerge in our universe. In this paper, we investigate two complex scalar fields, a baryonic scalar field and a leptonic one, in an Abelian gauge theory. We find stable solutions of gauged Q-balls for different baryon and lepton charges. Those solutions shows that a Coulomb potential arises and the Q-ball becomes electrically charged as expected. It is energetically favored that some amount of leptonic component decays, but there is an upper bound on its amount due to the Coulomb force. The baryonic decay also becomes possible by virtue of electrical repulsion and we find the condition to suppress it so that the charged Q-balls can survive in the universe.
\end{abstract}


\end{center}
\end{titlepage}

\baselineskip 6mm


\section{Introduction
\label{sec:introduction}}
Affleck-Dine mechanism~\cite{ad} is a promising candidate for baryogenesis in supersymmetric (SUSY) theories due to its consistency with the observational bound on reheating temperature which avoids the gravitino problem~\cite{kkrehgrav}. In the Affleck-Dine mechanism, baryon asymmetry is generated through the dynamics in the phase direction of a complex scalar field which carries nonzero baryon charge~\cite{gkmmssm}. The scalar field is called Affleck-Dine field. Affleck-Dine field is spatially homogeneous when it starts to oscillate, but spatial inhomogeneities due to quantum fluctuations grows exponentially into non-topological solitons, which are called Q-balls~\cite{ks,kkins,kkins2}. Q-ball is a spherical condensate of a scalar field and is defined as a solution in a global U(1) theory which minimizes energy of the system with its charge fixed~\cite{c}. Q-ball is known to decay into quarks or leptons, so that the final baryon number is carried by quarks produced through the decay of Q-balls. However, in gauge mediated SUSY breaking models, a baryonic Q-ball, with sufficiently large charge, can be stable against decay into nuclei when energy per charge of Q-ball is smaller than the proton mass~\cite{dk}. On the other hand, for a leptonic Q-ball, there exist decay channels into leptons. In this case, baryon number of the universe is generated from lepton asymmetry through sphaleron effect, i.e. leptogenesis. 

We focus on Q-balls that carry both baryon and lepton charges. In fact, these Q-balls can be formed when we consider the Affleck-Dine baryogenesis with $u^cu^cd^ce^c$ flat direction, for instance. In this case, it is possible that its lepton component can decay into leptons while its baryon component cannot decay into baryons. This implies that the difference between the baryonic component and the leptonic component in decay rate may induce an electric charge. Therefore, through the decay of the leptonic components only, an electric charge is induced even if the neutral Q-ball was formed at the beginning. The electric charge is expected to make differences in the experimental signatures of the relic Q-balls. For instance, the neutral Q-balls can be detected by Super-Kamiokande~\cite{sipal, kensyutu, kttn}, which probes the absorption of protons, but this detector is not suited for detection of the charged Q-balls since the charged Q-balls cannot absorb the protons due to the electrical repulsion. The charged Q-balls are likely to behave as some kind of nuclei, and are known to be detectable by such detectors as MACRO~\cite{sipal,macr} and the observational bounds on mass and flux of the relic charged Q-balls are obtained~\cite{kensyutu}.

In fact, electrically charged Q-ball has been studied in the literature and is called gauged Q-ball~\cite{gaugedqball}. However, the previous works studied gauged Q-balls in one scalar field theories, so that their results cannot be applied to Q-balls generated after the Affleck-Dine baryogenesis. There is a previous work which also discussed the evolution of Q-balls in two scalar model~\cite{ktt,sk}, in which the gauge field was neglected. However, in the course of decay, it is expected that an electrical repulsion arises, so that the gauge field must be taken into account.

In this paper, we consider a simplified model where there are two complex scalar fields and U(1) gauge field. One of the complex fields carries baryon charge and positive electric charge, while the other one carries lepton charge and negative electric charge. Our main purpose is to demonstrate that gauged Q-balls may be realized in our universe through the decay process of the lepton component, even if Q-balls are initially neutral. We find stable solutions for different baryon and lepton charges, taking into account the effect of the gauge field. If we suppose that only the leptonic component decays off, a sequence of solutions with $B=const.$ represents the decay process. We examine quantitatively whether such a process is energetically allowed.

In the following sections, we first review the main properties of global Q-balls and gauged Q-balls, i.e. neutral and charged Q-balls respectively. Subsequently, in order to discuss the evolution of a Q-ball which is formed from the flat direction, we consider a two scalar model and find sequences of gauged Q-ball solutions with $B=const$. We calculate the total energy of Q-balls and gauge fields and examine whether the leptonic decay is energetically favorable. Finally, we give our conclusions in Sec.~\ref{sec:conc}.
\section{Global Q-ball}
\label{sec:q-ball}
In this section, we briefly review the main properties of a global Q-ball, which is a stable configuration of a complex scalar field with a fixed conserved charge. Consider a theory of a complex scalar field with a global U(1) charge. The Lagrangian is written as
\begin{align}
\mathcal{L}=\partial_\mu\Phi^\ast\partial^\mu\Phi-V(\Phi)
\end{align}
where $V(\Phi)$ is a scalar potential and is normalized so that $V(0)=0$. Let us renormalize the field for later convenience as
\begin{align}
\label{eq:scred}
\Phi\equiv\frac1{\sqrt2}\phi.
\end{align}
The global U(1) charge density $q$ is given by 
\begin{align}
q=\frac1{2i}\left(\phi^{\ast}\dot{\phi}-\phi\dot{\phi^{\ast}}\right).
\end{align}
Q-ball is defined as a solution which minimizes energy of the system with its charge fixed. Using the Lagrange multiplier method,  we only need to minimize the following function: 
\begin{align}
\label{eq:energy}
E_{\omega}\equiv E+\omega\left[Q-\frac1{2i}\int d^3x(\phi^\ast\dot\phi-\phi\dot\phi^\ast)\right]
\end{align}
where $E$ is given by
\begin{align} 
E=\int{d^3x\left[\frac12\left(\vert \dot{\phi} \vert^2+\left\vert\nabla\phi\right\vert^2\right)+V(\phi)\right]}.
\end{align}
Equation~(\ref{eq:energy}) is rewritten as
\begin{align}
\label{eq:en2}
&E_{\omega}=\int d^3x\frac12|\partial_t\phi-i\omega\phi|^2+\int d^3x\left[\frac12|\nabla\phi|^2+V_\omega(\phi)\right]+\omega Q,\\
&V_\omega(\phi)\equiv V(\phi)-\frac12\omega^2|\phi|^2.
\end{align}
By minimizing the first term of Eq.~(\ref{eq:en2}), we can derive the time dependence of the solution as
\begin{align}
\label{eq:qpara}
\phi(x, t)=e^{i\omega t}\phi(x).
\end{align}
Moreover, it is known that the solution which minimizes the second term of Eq.~(\ref{eq:en2}) is real and spatially symmetric~\cite{ksph}. 
Therefore, the radial direction $\phi(r)$ is a solution of the following equation: 
\begin{align}
\label{eq:eqmq}
\frac{d^2\phi}{dr^2}+\frac2r\frac{d\phi}{dr}+\left[\omega^2\phi(r)-\frac{\partial V(\phi)}{\partial\phi}\right]=0.
\end{align}
Here, in order to avoid a singularity at $r=0$ and to find a specially localized solution, we set boundary conditions as
\begin{align}
\label{eq:bc}
\frac{d\phi}{dr}(0)=0,~~ \phi(\infty)=0.
\end{align}

\begin{figure}[t]
\begin{center}
  \includegraphics[width=0.65\linewidth]{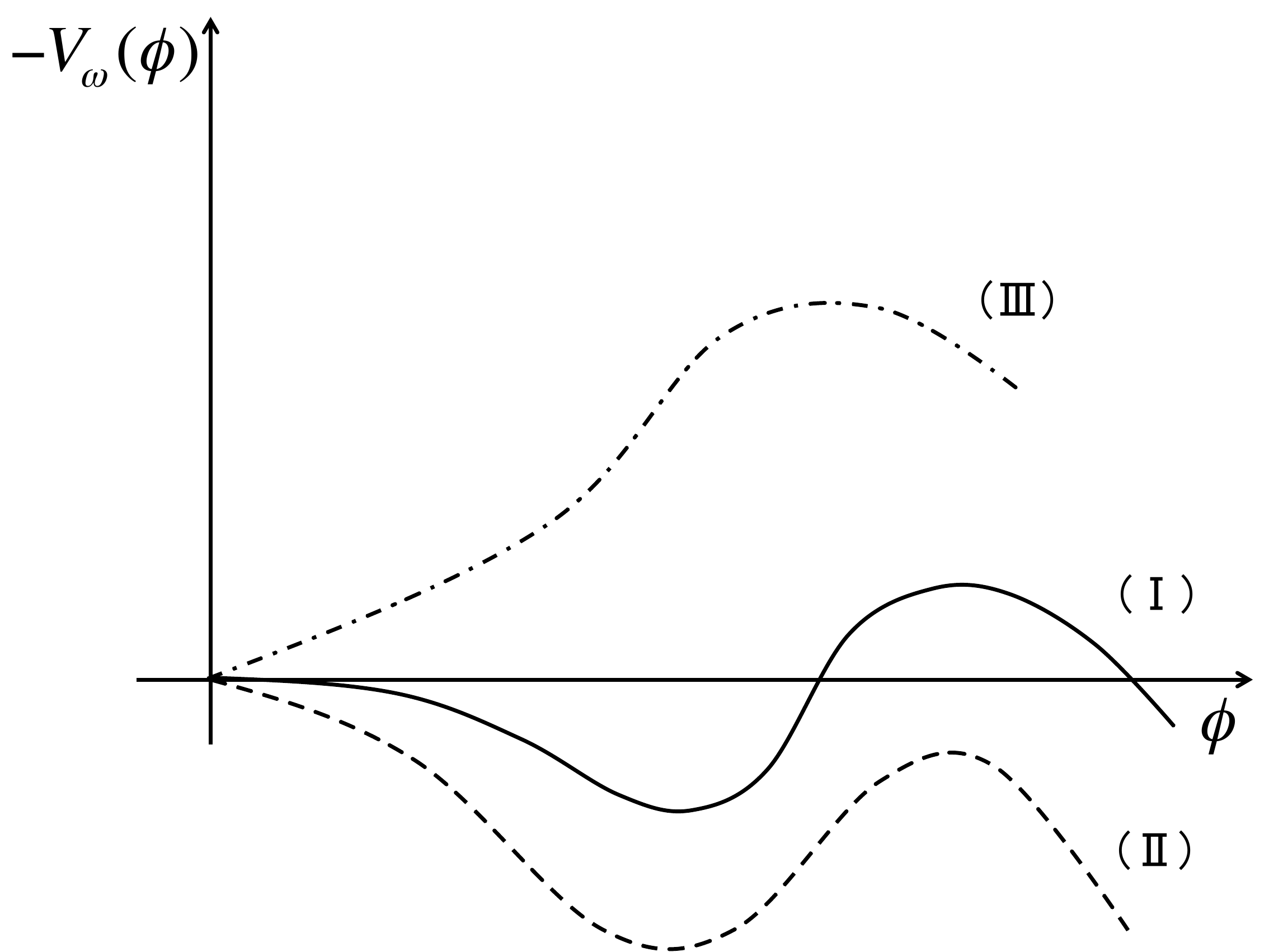}
 \caption{Several kinds of shape of $-V_\omega(\phi)$ depending on $\omega$. (I)\ :\ $\omega_0^2<\omega^2<V''(0)$,\ (II)\ :\ $\omega^2<\omega_0^2$,\ (III)\ :\ $\omega^2>V''(0)$. The potential of type (II) is appropriate for the existence of the solution.}
\label{fig:syuqball}
\end{center}
\end{figure}

Now let us find the condition that there exists a solution of Eq.~(\ref{eq:eqmq}). First, we redefine the spatial coordinate and the potential as 
\begin{align}
\label{eq:redef}
r&\to t,\\
-V_\omega(\phi)&=\frac12\omega^2\phi^2-V(\phi).
\end{align}
The equation of motion then becomes 
\begin{align}
&\ddot\phi+\frac2t\dot\phi-\frac{\partial V_\omega}{\partial\phi}=0
\end{align}
and we see that it is equivalent to the equation of motion of a classical point particle under the potential $-V_\omega$. The boundary conditions Eq.~(\ref{eq:bc}) imply that the classical particle is initially at rest and converges toward the origin asymptotically. In order that this kind of motion is possible, the potential $-V_\omega$ should be in an appropriate shape as shown in Fig.~\ref{fig:syuqball}.
Quantitatively, $-V_\omega$ must satisfy the following conditions:
\begin{align}
\max[-V_\omega(\phi)]>0,
\end{align}
\begin{align}
-V_\omega''(0)<0.
\end{align}
The first condition is necessary in order for the particle to possess the potential energy enough to reach the origin. The second means that the particle must be subjected to a backward force near the origin and is also necessary in order for the particle to stop at the origin, not to roll over it.
The conditions above are rewritten as the conditions to $\omega$ as
\begin{align}
\label{eq:omc}
&\omega_0^2<\omega^2<V''(0),\\
&\omega_0^2\equiv \min\left[\frac{2V(\phi)}{\phi^2}\right].
\end{align}

Furthermore, it generally holds that 
\begin{align}
\frac{dE}{dQ}=\omega, 
\end{align}
for any Q-ball solution, which can be easily shown by taking a variation of energy in the following way:
\begin{align}
\delta E&=\int{d^3x\left[\omega\delta\omega\phi^2+\omega^2\phi\delta\phi-\Delta\phi\delta\phi+V'(\phi)\delta\phi\right]}\nonumber\\
&=\omega\int{d^3x\left[\delta\omega\phi^2+2\omega\phi\delta\phi\right]}\nonumber\\
&=\omega\delta Q,
\end{align}
where we used the equation of motion Eq.~(\ref{eq:eqmq}).
Therefore, the condition Eq.~(\ref{eq:omc}) is rewritten as
\begin{align}
&\omega_0<\frac{dE}{dQ}<\sqrt{V''(0)}\equiv m_\Phi.
\end{align} 
It is noted that the second inequality indicates that a Q-ball with a charge $Q$ is energetically favorable compared to a Q-ball with a charge $Q-1$ and a particle, which is consistent with the definition of Q-ball solution. 

Here, let us identify the scalar field $\phi$ as a D- and F- flat direction in gauge mediated SUSY breaking models. In this case, the potential of the flat direction is mainly given by its soft mass term. However, the soft mass is suppressed for energy scale larger than messenger scale and the potential becomes flat\footnote{The exact form of the potential is derived in Ref.~\cite{mmgc}.}. If we approximate the potential as $V=V_0=const.$, there exists an analytic solution~\cite{dk} given by
\begin{align}
\phi=\left\{\begin{array}{ll}\phi_0\sin(\omega r)/\omega r,&r<R\equiv\pi/\omega\\[0.6cm]
0,&r>R\end{array}\right.\\
\end{align} 
whose energy is then written as
\begin{align}
E=\frac{4\pi\sqrt{2}}3V_0^{1/4}Q^{3/4}.
\end{align}
Then, from $dE/dQ=\omega$, which is proven above, 
\begin{align}
\omega=\sqrt{2}\pi V_0^{1/4}Q^{-1/4}.
\end{align}
This indicates that for a large charge, $dE/dQ$ may become small enough. Therefore, for a baryonic Q-ball with a large baryon number, it may hold that
\begin{align}
\frac{dE}{dB}<m_p,
\end{align}  
where $m_p$ ($\simeq1\mathrm{GeV}$) is the proton mass. This means that the Q-ball is stable against the decay into protons. On the other hand, if Q-ball carries a lepton number, it can decay into leptons. This is the motivation of our assumption in Sec.~\ref{sec:qtwo} that baryonic component of Q-ball is stable while leptonic component can decay into leptons.
\section{Gauged Q-ball}
\label{sec:gauged q-ball}
In this section, we consider a complex scalar field that is charged under U(1) gauge symmetry. Although this toy model is not motivated in SUSY theories, we investigate it to see the effect of gauge force on Q-balls. In this case, we must solve the equation of motion for the gauge field $A_\mu$ as well as the complex scalar field. The spatially localized configuration in this theory is called gauged Q-ball.  

The Lagrangian is written as 
\begin{align}
\mathcal{L}&=(D_\mu\Phi)^\ast D^\mu\Phi-V(\Phi)-\frac14F_{\mu\nu}F^{\mu\nu},\\
D_\mu&\equiv\partial_\mu-ieA_\mu
\end{align}
where $V(\Phi)$ is a scalar potential. We parameterize the scalar field in the same way as in the case of global Q-ball:
\begin{align}
&\Phi\equiv\frac1{\sqrt2}\phi,\\
\label{eq:gqpara}
&\phi(x,t)=e^{i\omega t}\phi(r).
\end{align}
For the gauge field, we adopt the following parameterization~\cite{gaugedqball}.
\begin{align}
&A_0=A_0(r),\\
&A_i=0.
\end{align}
The first indicates that we are searching the spatially symmetric solution and the second implies that we are assuming the absence of a magnetic field, which means in turn the absence of an electric current. The equations of motion are then given by
\begin{align}
\label{eq:eomf}
&\frac{d^2\phi}{dr^2}+\frac2r\frac{d\phi}{dr}+\phi g^2-\frac{dV}{d\phi}=0,\\
&\frac{d^2g}{dr^2}+\frac2r\frac{dg}{dr}-e^2\phi^2g=0,
\label{eq:eomgaugedqball}
\end{align}
where we redefine the gauge field to absorb $\omega$ as $g\equiv\omega-eA_0$. Note that $g$ is gauge invariant. We set boundary conditions as 
\begin{align}
\label{eq:bdcga}
&\phi(\infty)=0,~~ \frac{d\phi}{dr}(0)=0,\\
&g(\infty)=\omega,~~\frac{dg}{dr}(0)=0,
\end{align}
to avoid singularities at $r=0$. Note that when $\phi(r)\to0$ as $r\to\infty$, the gauge field asymptotes to a certain constant as $r\to\infty$ by Eq.~(\ref{eq:eomgaugedqball}).
Thus the boundary condition $g(\infty)=\omega$ is just a definition of $\omega$.
Note that Eq.~(\ref{eq:eomgaugedqball}) can be rewritten as 
\begin{align}
(r^2g')'=e^2r^2\phi^2g.
\end{align}
This implies that if $g(0)>0$, then $g'$ becomes positive for $r>0$, so that $g$ increases, while in the opposite case $g(0)<0$, $g$ decreases. In either case, $g^2$ always increases. 

\begin{figure}[t]
\begin{minipage}{.45\linewidth}
  \includegraphics[width=\linewidth]{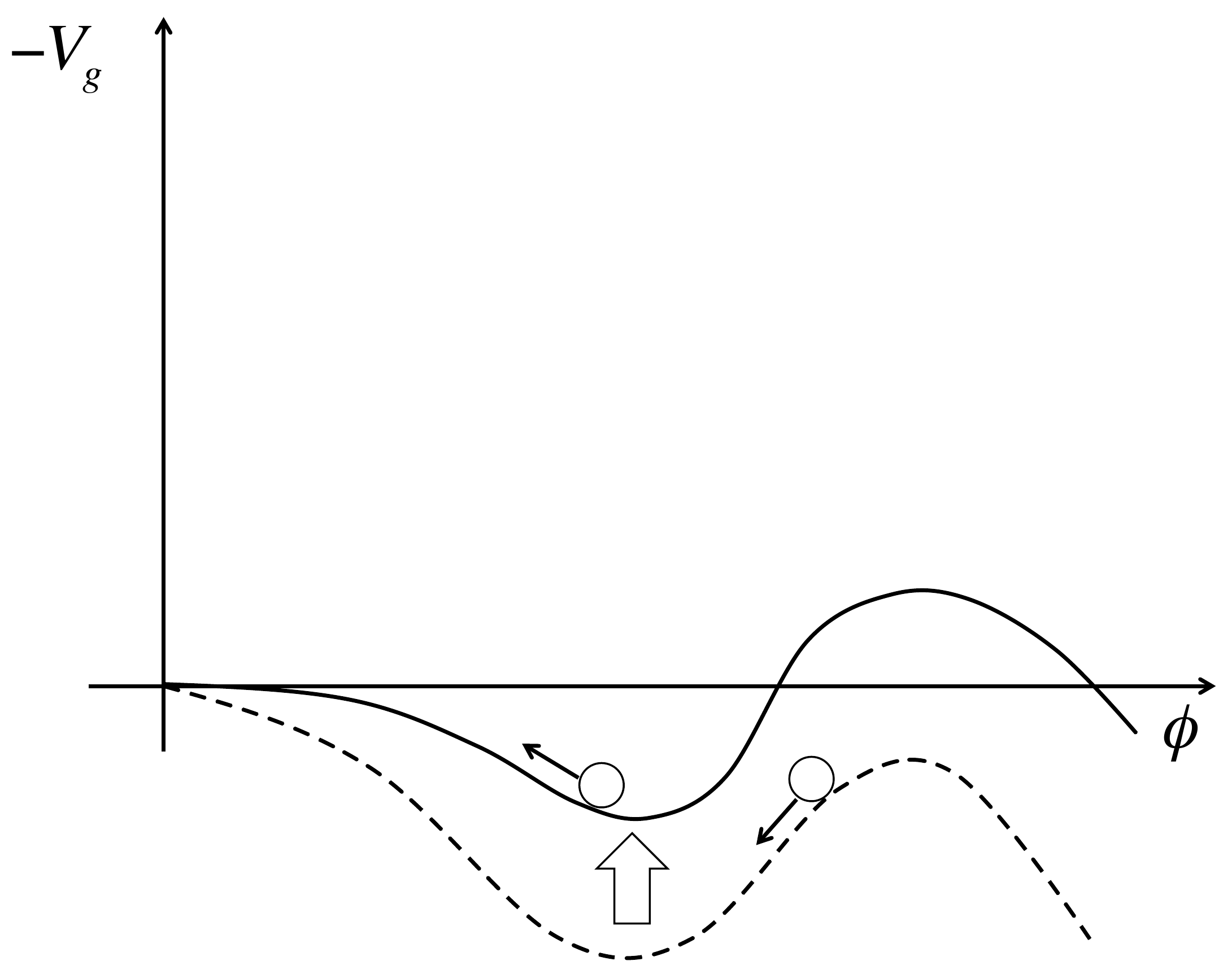}
 \end{minipage}
 \hspace{1cm}
 \begin{minipage}{.45\linewidth}
  \includegraphics[width=\linewidth]{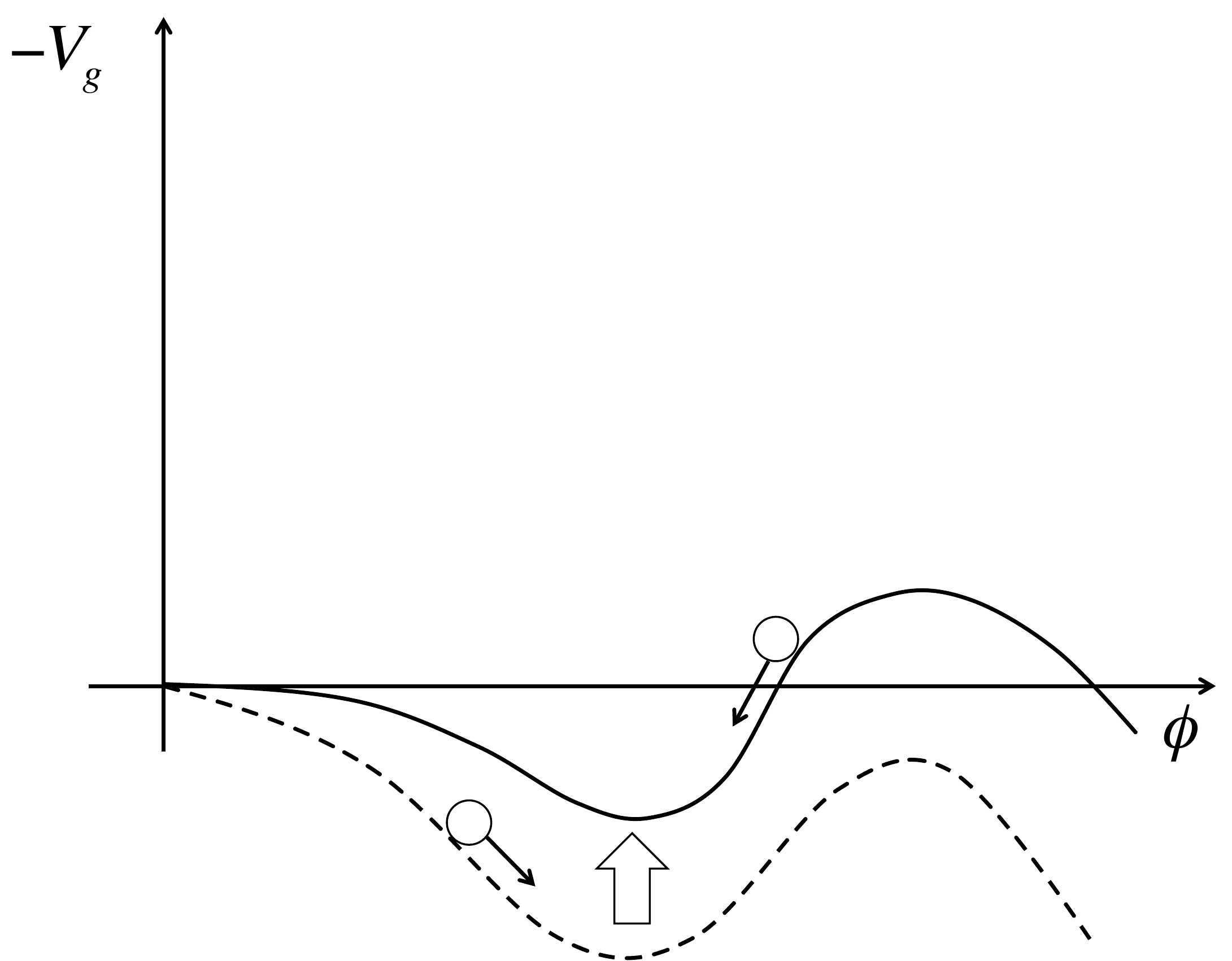}
 \end{minipage}
 \caption{Mechanical analogy of gauged Q-ball}
\label{fig:gaugedpoten2}
\end{figure}

\begin{figure}[t]
\begin{minipage}{.45\linewidth}
  \includegraphics[width=\linewidth]{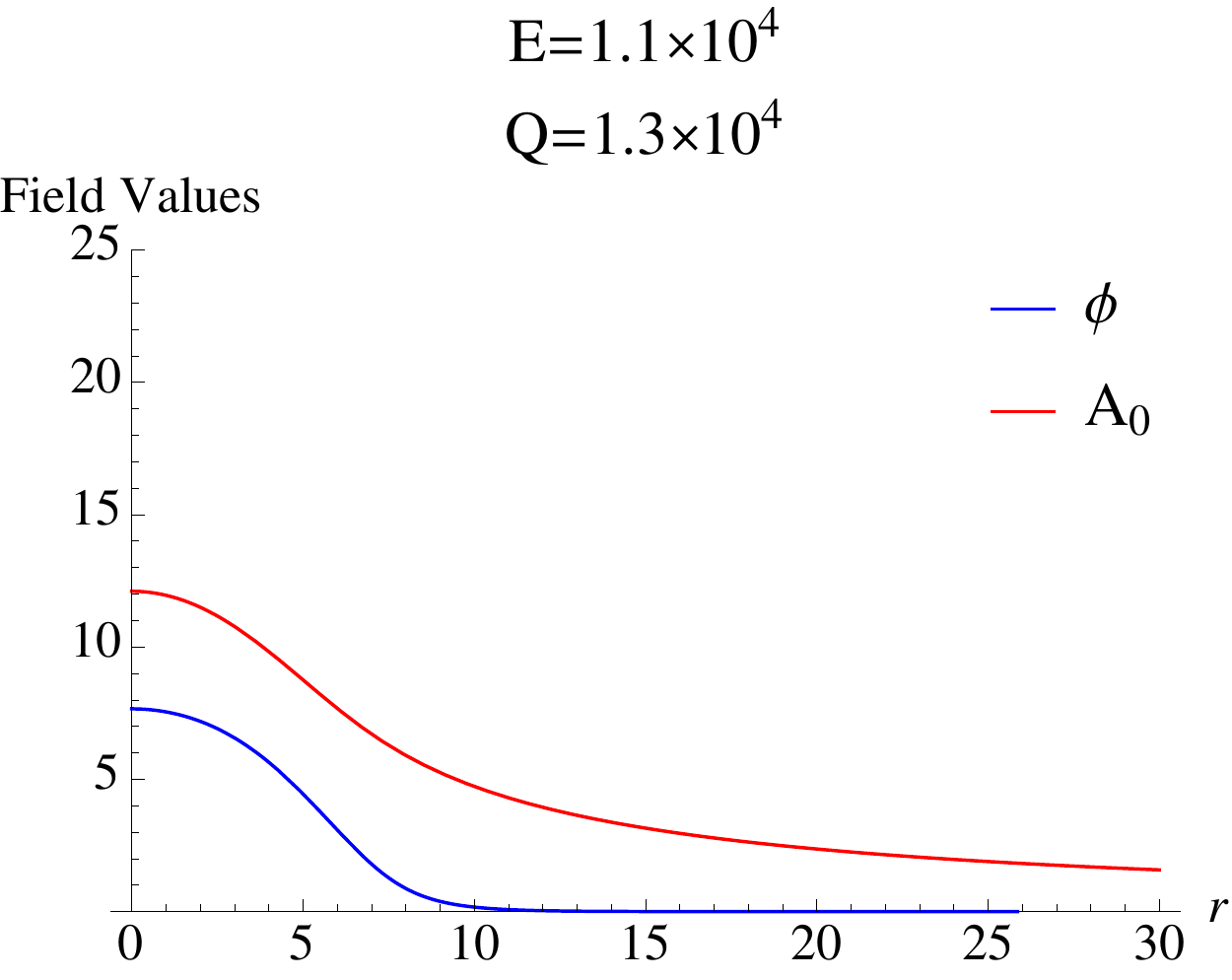}
 \end{minipage}
 \hspace{1cm}
 \begin{minipage}{.45\linewidth}
  \includegraphics[width=\linewidth]{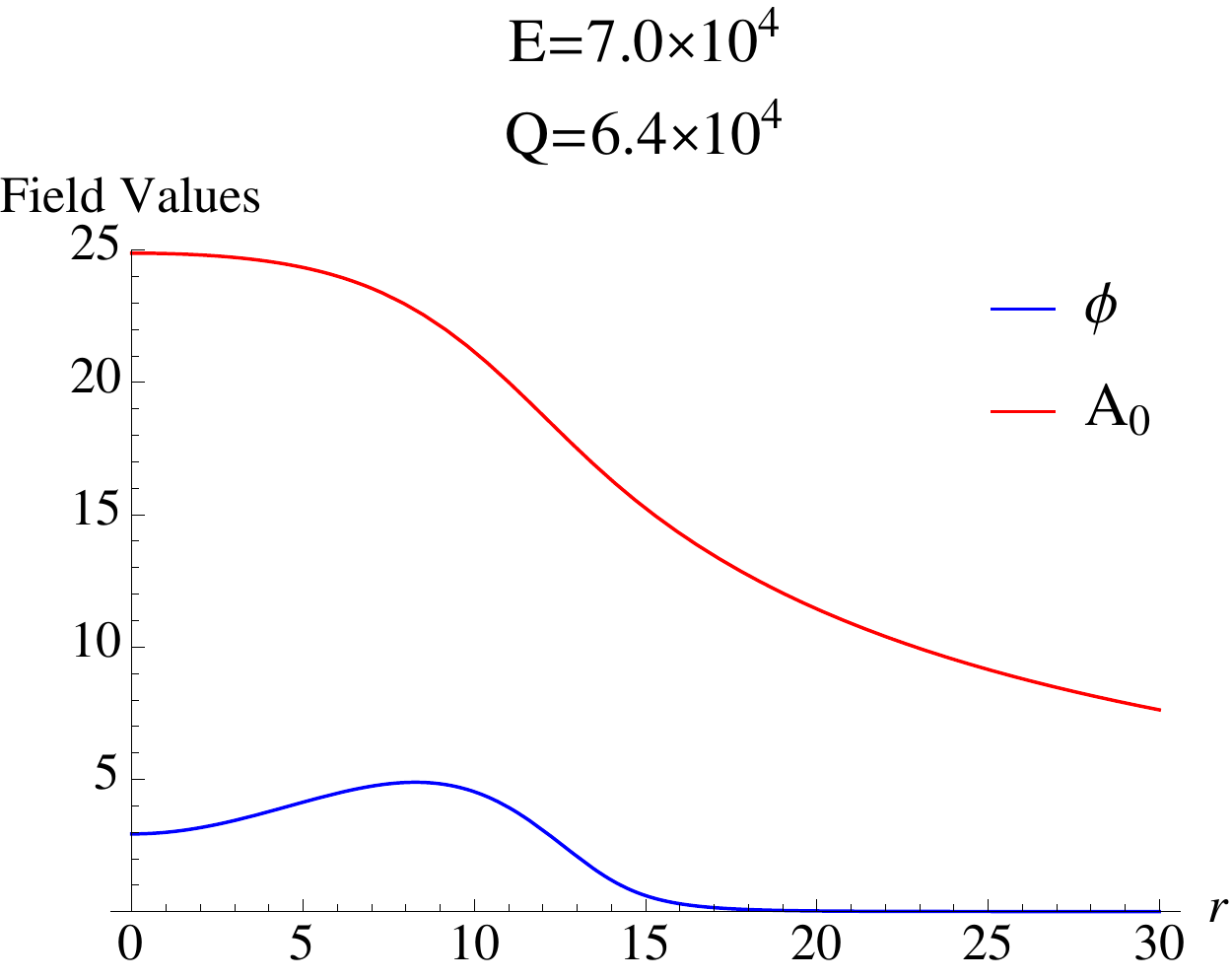}
 \end{minipage}
 \caption{Gauged Q-balls in gauge mediation-like model with $e^2=0.002$ and $m_\Phi=1$.}
\label{fig:gauex}
\end{figure}

Here, let us consider an analogy in a similar way as considered in the previous section (Fig.~\ref{fig:gaugedpoten2}). Eq.~(\ref{eq:eomf}) is analogous to the equation of motion for a classical particle which is subjected to the potential $-V_g=-V+\frac12g^2\phi^2$. As mentioned above, $g^2$ always increases, so that the effective mass term $g^2\phi^2/2$ increases with time. From this we can derive some important properties of gauged Q-ball. Since the effective mass increases, it is energetically possible that the particle reaches the origin even if it moves away from the origin at the beginning, which means that there exists radially non-monotonic solution. We can interpret this kind of solution as the result of the scalar field being pushed outward due to the electrical repulsion. We show both kinds of solutions for gauge mediation-like model in Fig.~\ref{fig:gauex}, where we approximate the potential as $V(\Phi)=m_\Phi^4\ln(1+\Phi^2/m_\Phi^2)$. Indeed the non-monotonic solutions arise for charges larger than those of the monotonic ones.  
\begin{figure}[t]
\begin{center}
  \includegraphics[width=0.7\linewidth]{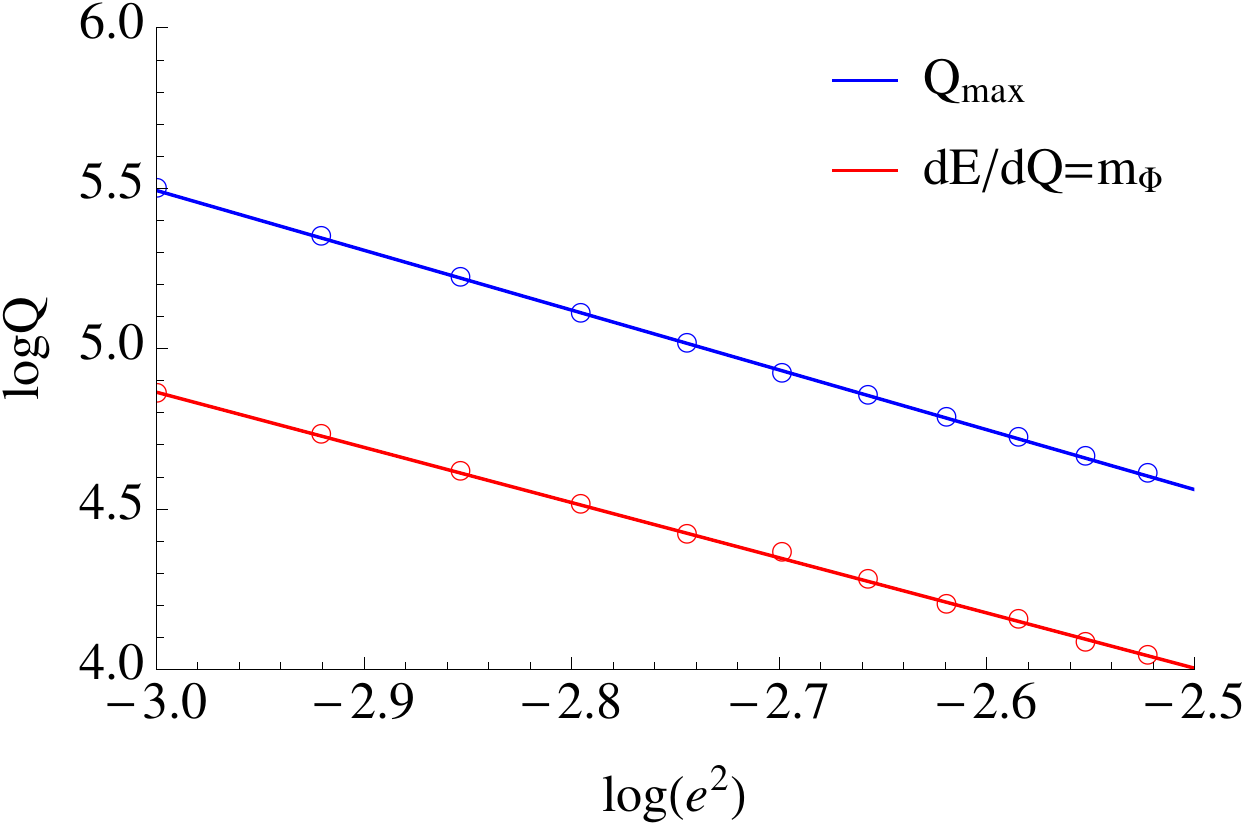}
 \caption{Illustration of upper limits on $Q$, and where $dE/dQ=m_\Phi$}
\label{fig:1sdedq1}
\end{center}
\end{figure}

It holds that 
\begin{align}
\frac{dE}{dQ}=\omega,
\end{align}
for gauged Q-ball as well, whose proof is similar to that for global Q-ball~\cite{deodq}. The energy of the Q-ball is given by 
\begin{align}
E&=\int d^3x\left[\frac12(\nabla\phi)^2+\frac1{2}(\nabla A_0)^2+\frac12\phi^2(\omega-eA_0)^2+V(\phi)\right],
\end{align}
whose variation leads to
\begin{align}
\delta E&=\int d^3x\left[-\Delta\phi\delta\phi-A_0\delta\Delta A_0+\phi\delta\phi (\omega-eA_0)^2 +(\omega-eA_0)\delta (\omega-eA_0)\phi^2+V'\delta\phi\right]\nonumber\\
&=\int d^3x\left[2\phi\delta\phi (\omega-eA_0)^2+(\omega-eA_0)\delta (\omega-eA_0)\phi^2-A_0\delta\Delta A_0\right]\nonumber\\
&=\int d^3x\left[(\omega-eA_0)\delta q-A_0\delta\Delta A_0\right]\nonumber\\
&=\omega\delta Q-\int d^3x\left[eA_0\delta q+A_0\delta\Delta A_0\right]\nonumber\\
&=\omega\delta Q.
\end{align}
Here we used Eqs.~(\ref{eq:eomf}) and (\ref{eq:eomgaugedqball}), and the charge of the Q-ball is given by
\begin{align}
Q&\equiv \int q d^3 x \\
&=\int d^3 x(\omega-eA_0)\phi^2.
\end{align}

The charge of a gauged Q-ball has an upper limit, above which there is no localized solution. This is contrast to the case of global Q-balls, where Q-ball solutions exist for arbitrarily large $Q$. The blue circles in Fig.~\ref{fig:1sdedq1} are the upper limits on charges, and the results can be fitted as 
\begin{align}
\log_{10} Q_{\rm max}=-1.0\times10^{-1}-1.9\log_{10}(e^2),
\end{align}
which is shown by the blue line.

For a global Q-ball, the condition
\begin{align}
\frac{dE}{dQ}<m_\Phi,
\end{align} 
was necessary for the existence of solution (see Eq.~(\ref{eq:omc})). However, for a gauged Q-ball, it is possible that
\begin{align}
\label{eq:metas}
\frac{dE}{dQ}>m_\Phi,
\end{align}
which implies a Q-ball with a charge $Q-1$ and a particle at infinity are energetically favorable compared to a Q-ball with a charge $Q$. This behavior arises when the charge is large enough and therefore may be interpreted as a result of electrical repulsion. Even so, a gauged Q-ball may exist as itself because a Q-ball with a charge $Q$ is still energetically favorable compared to a Q-ball with a charge $Q-1$ and a particle near the surface just after emission, due to the large Coulomb potential. Thus, gauged Q-ball is expected to be a metastable solution if Eq.~(\ref{eq:metas}) is satisfied. 
The red circles in Fig.~\ref{fig:1sdedq1} demonstrates when $dE/dQ$ becomes $m_\Phi$. The results can be fitted as 
\begin{align}
\log_{10} Q=-1.1\times10^{-1}-1.6\log_{10}(e^2).
\end{align} 
This dependence can be explained in the following way.
If we approximate the energy of an emitted particle $dE/dQ$ as the energy with electricity switched off, plus Coulomb energy, 
\begin{align}
\frac{dE}{dQ}&=\omega_0+\frac{e^2Q}{4\pi R}\nonumber\\
&=\omega_0+\frac{e^2Q}{4\pi^2}\omega_0\nonumber\\
&=\omega_0+\frac{e^2Q}{2\sqrt{2}\pi}V_0^{1/4}Q^{-1/4},   
\end{align}
where we used the analytic expression in the previous section with $R=\pi/\omega_0$, $\omega_0=\sqrt2\pi V_0^{1/4}Q^{-1/4}$.
The charge of Q-ball at which $dE/dQ=m_\Phi$ is thus given by
\begin{align}
Q=\left(\frac{2\sqrt{2}\pi m_\Phi}{e^2V_0^{1/4}}\right)^{4/3},
\end{align}
or
\begin{align}
\log_{10} Q&=\frac43\log_{10}\left(\frac{2\sqrt{2}\pi m_\Phi}{V_0^{1/4}}\right)-\frac43\log_{10}(e^2)\nonumber\\
&\simeq9.9\times10^{-1}-1.3\log_{10}(e^2),
\end{align}
where we used $m_\Phi\gg\omega_0$ and set $V_0$ into $V_0^{1/4}=1.6m_\Phi$, whose value is appropriate for the solutions we are dealing with. This estimation roughly explains our numerical solutions of Fig.~\ref{fig:1sdedq1}. 
\section{Q-balls in two scalar model with U(1) gauge field}
\label{sec:qtwo}
Since our main interest is in the evolution of global Q-balls formed from the flat direction which is responsible for the Affleck-Dine mechanism, we must consider the case of several numbers of scalar fields coupled with a gauge field. Here we consider the simplest case in which the flat direction consists of two scalar fields which carry baryon and lepton numbers respectively, and the gauge field is Abelian. The electric charges must be opposite since the flat direction is neutral. The Lagrangian is then written as
\begin{align}
&\mathcal{L}=(D_\mu\Phi_1)^\ast D^\mu\Phi_1+(D_\mu\Phi_2)^\ast D^\mu\Phi_2-V(\Phi_1,\Phi_2)-\frac14F_{\mu\nu}F^{\mu\nu},\\
&D_\mu\Phi_1=(\partial_\mu-ieA_\mu)\Phi_1,\\
&D_\mu\Phi_2=(\partial_\mu+ieA_\mu)\Phi_2
\end{align}
and baryon and lepton charges are
\begin{align}
&B=\frac1{i}\int d^3x(\Phi_1^\ast D_0\Phi_1-\Phi_1(D_0\Phi_1)^\ast)\equiv\int d^3xb,\\
&L=\frac1{i}\int d^3x(\Phi_2^\ast D_0\Phi_2-\Phi_2(D_0\Phi_2)^\ast)\equiv\int d^3xl
\end{align}
where $b$ and $l$ are baryon and lepton number densities.
Since we assign the positive and negative electric charges 
for B and L components, respectively, 
the total electric charge is given by 
\begin{align}
Q=B-L.
\end{align}

We find stable solutions and calculate their energies, and examine if leptonic decay is energetically allowed. 
First, we adopt the same parameterization as before:
\begin{align}
\Phi_i&\equiv\frac1{\sqrt2}\phi_i,\ \ \   i=1,2\\
\phi_1(x,t)&=e^{i\omega_1t}\phi_1(r),\\
\phi_2(x,t)&=e^{i\omega_2t}\phi_2(r),\\
A_i&=0,\\
A_0&=A_0(r).
\end{align}
The equations of motion then become 
\begin{align}
\label{eq:eom2s}
&\frac{d^2\phi_1}{dr^2}+\frac2r\frac{d\phi_1}{dr}+\phi_1(\omega_1-eA_0)^2-\frac{\partial V}{\partial\phi_1}=0,\\
&\frac{d^2\phi_2}{dr^2}+\frac2r\frac{d\phi_2}{dr}+\phi_2(\omega_2+eA_0)^2-\frac{\partial V}{\partial\phi_2}=0,\label{eq:e25}\\
&\frac{d^2A_0}{dr^2}+\frac2r\frac{dA_0}{dr}+e\phi_1^2(\omega_1-eA_0)-e\phi_2^2(\omega_2+eA_0)=0
\label{eq:eom2s3}
\end{align}
with the boundary conditions given by
\begin{align}
\phi_1(\infty)&=\phi_2(\infty)=0,\\
\frac{d\phi_1}{dr}(0)&=\frac{d\phi_2}{dr}(0)=0,\\
A_0(\infty)&=\frac{dA_0}{dr}(0)=0.
\end{align}
We prove here 
\begin{align}
&\left(\frac{\del E}{\del B}\right)_L=\omega_1,\\
&\left(\frac{\del E}{\del L}\right)_B=\omega_2
\end{align}
for later use, which is analogous to $dE/dQ=\omega$ for 1-scalar gauged Q-ball. The energy of the system is
\begin{align}
E=\int d^3x\left[\frac12(\nabla\phi_1)^2+\frac12(\nabla\phi_2)^2+\frac1{2e^2}(e\nabla A_0)^2+\frac12\phi_1^2(\omega_1-eA_0)^2+\frac12\phi_2^2(\omega_2+eA_0)^2+V(\phi_1,\phi_2)\right],
\end{align}
and its variation with respect to $\phi_1$, $\phi_2$ and $A_0$ is given by 
\begin{align}
\delta E&=\int d^3x\left[-\Delta\phi_1\delta\phi_1-\Delta\phi_2\delta\phi_2-A_0\delta\Delta A_0+\frac{\del V}{\del\phi_1}\delta\phi_1+\frac{\del V}{\del\phi_2}\delta\phi_2\right]\nonumber\\
&+\int d^3x\left[\phi_1\delta\phi_1(\omega_1-eA_0)^2+\phi_1^2(\omega_1-eA_0)\delta(\omega_1-eA_0)\right]\nonumber\\
&+\int d^3x\left[\phi_2\delta\phi_2(\omega_2+eA_0)^2+\phi_2^2(\omega_2+eA_0)\delta(\omega_2+eA_0)\right]\nonumber\\
&=\int d^3x\left[(\omega_1-eA_0)\delta b+(\omega_2+eA_0)\delta l-A_0\delta\Delta A_0\right]\nonumber\\
&=\omega_1\delta B+\omega_2\delta L+\int d^3xA_0\left[e(-\delta b+\delta l)-\delta\Delta A_0\right]\nonumber\\
&=\omega_1\delta B+\omega_2\delta L,
\end{align}
where we used Eqs.~(\ref{eq:eom2s}), (\ref{eq:e25}), and (\ref{eq:eom2s3}). The proof can be easily generalized into the case of arbitrary number of scalar fields.

\begin{figure}[!t]
\begin{minipage}{.45\linewidth}
  \includegraphics[width=\linewidth]{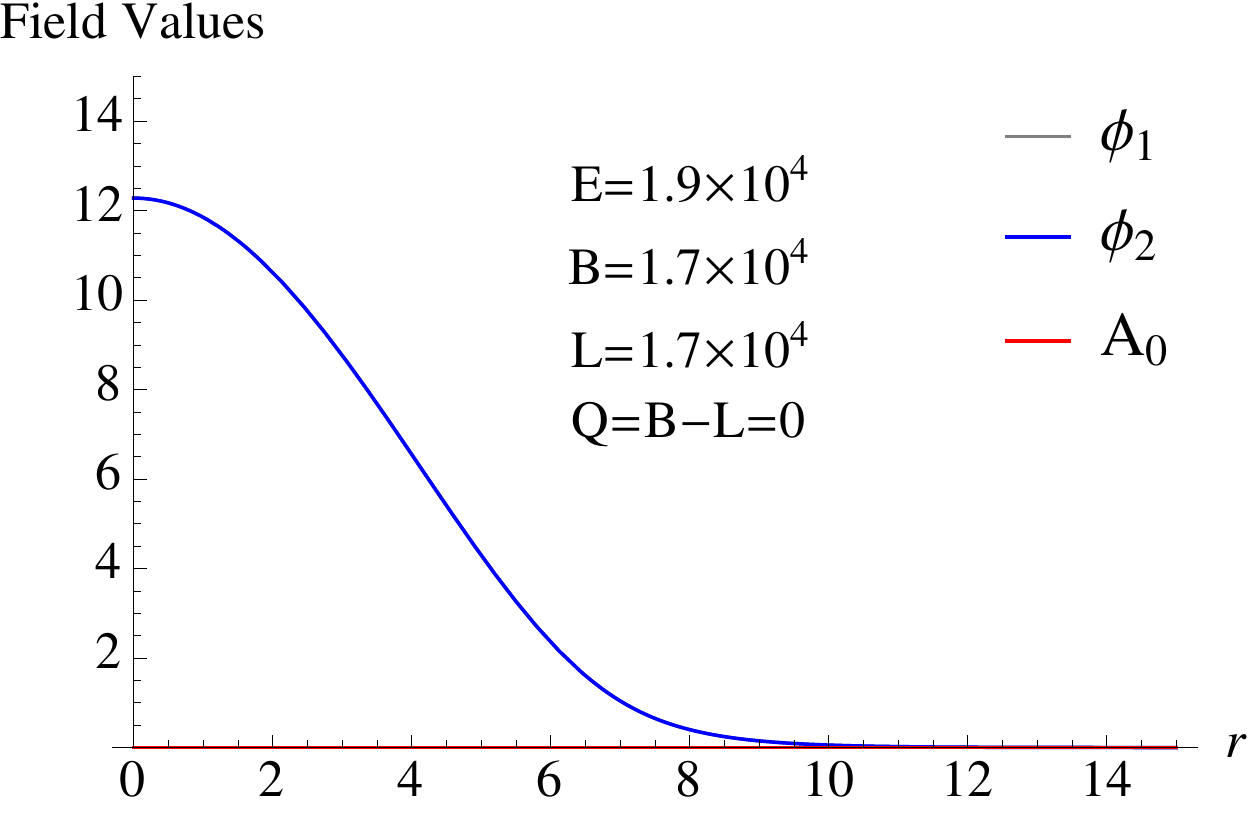}
 \end{minipage}
 \hspace{1cm}
 \begin{minipage}{.45\linewidth}
  \includegraphics[width=\linewidth]{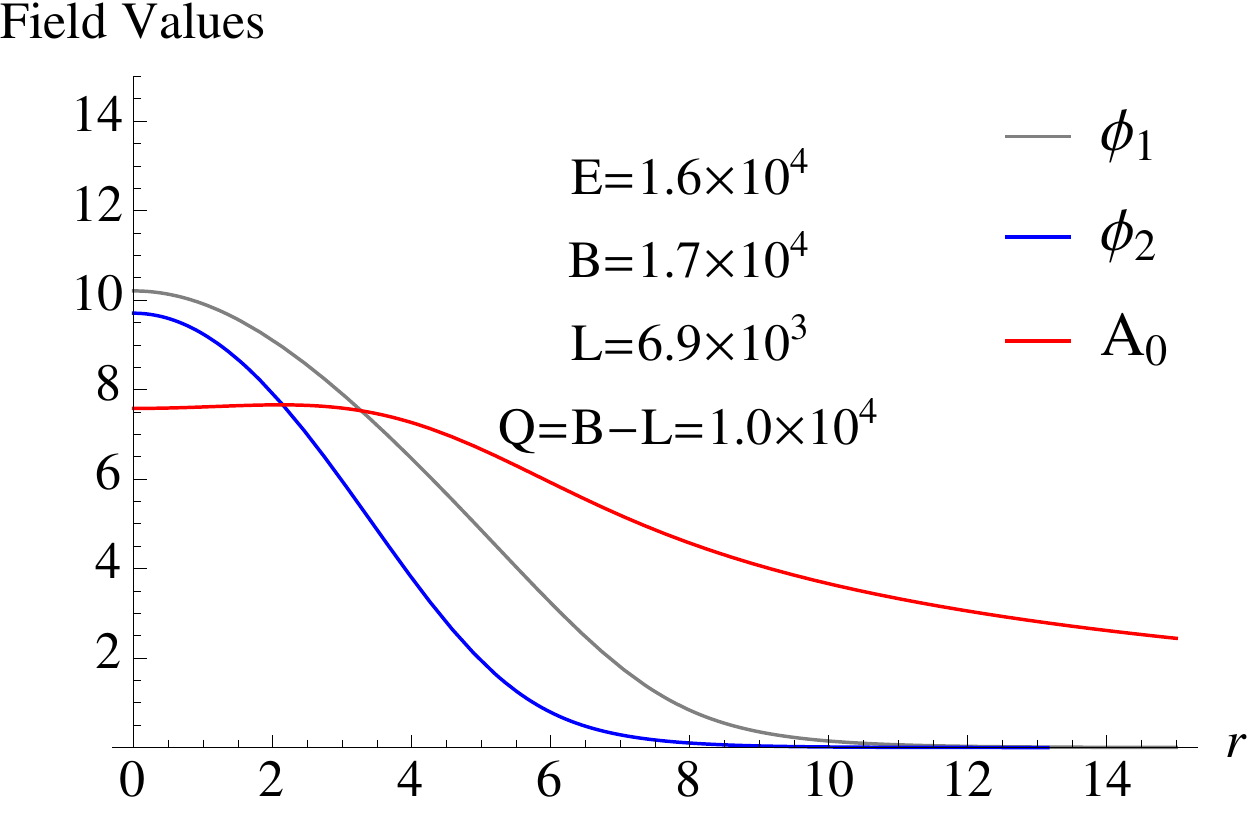}
 \end{minipage}
 \begin{minipage}{.45\linewidth}
  \includegraphics[width=\linewidth]{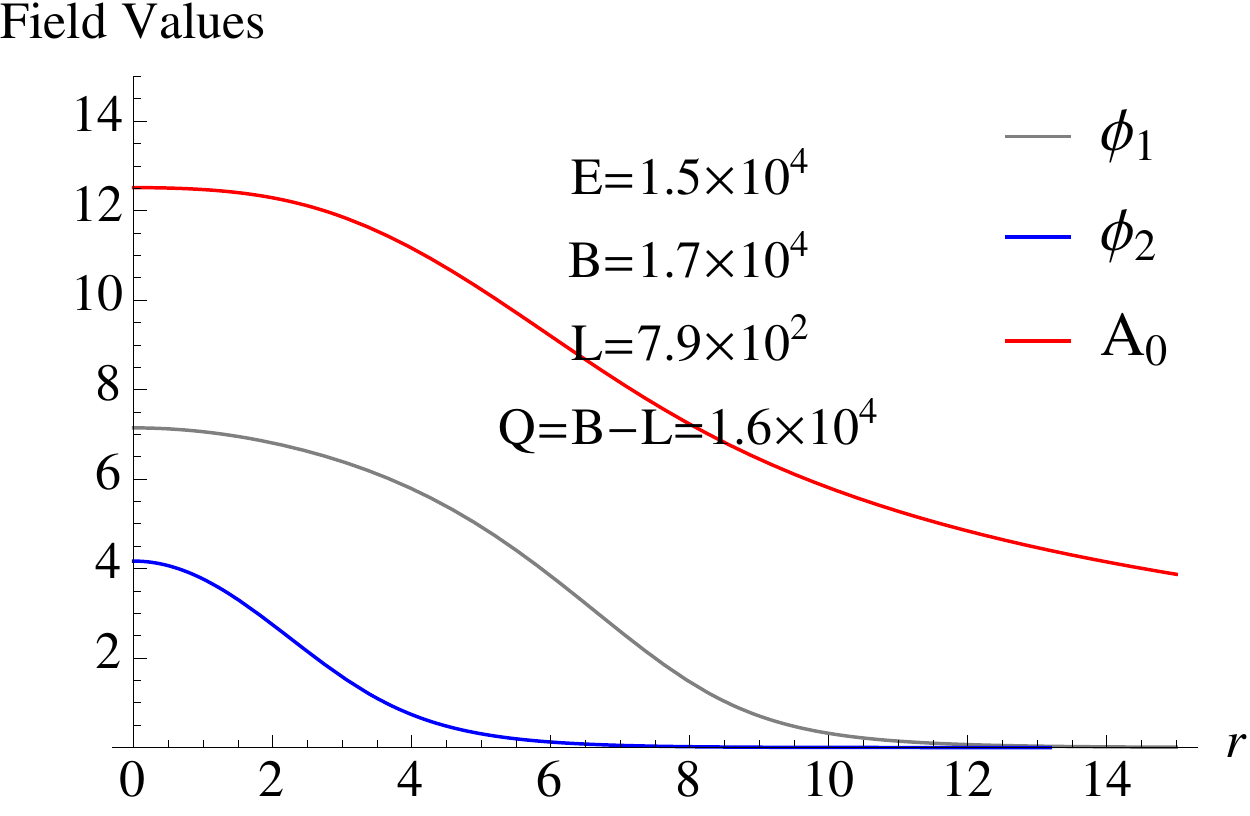}
 \end{minipage}
 \hspace{1.4cm}
 \begin{minipage}{.45\linewidth}
  \includegraphics[width=\linewidth]{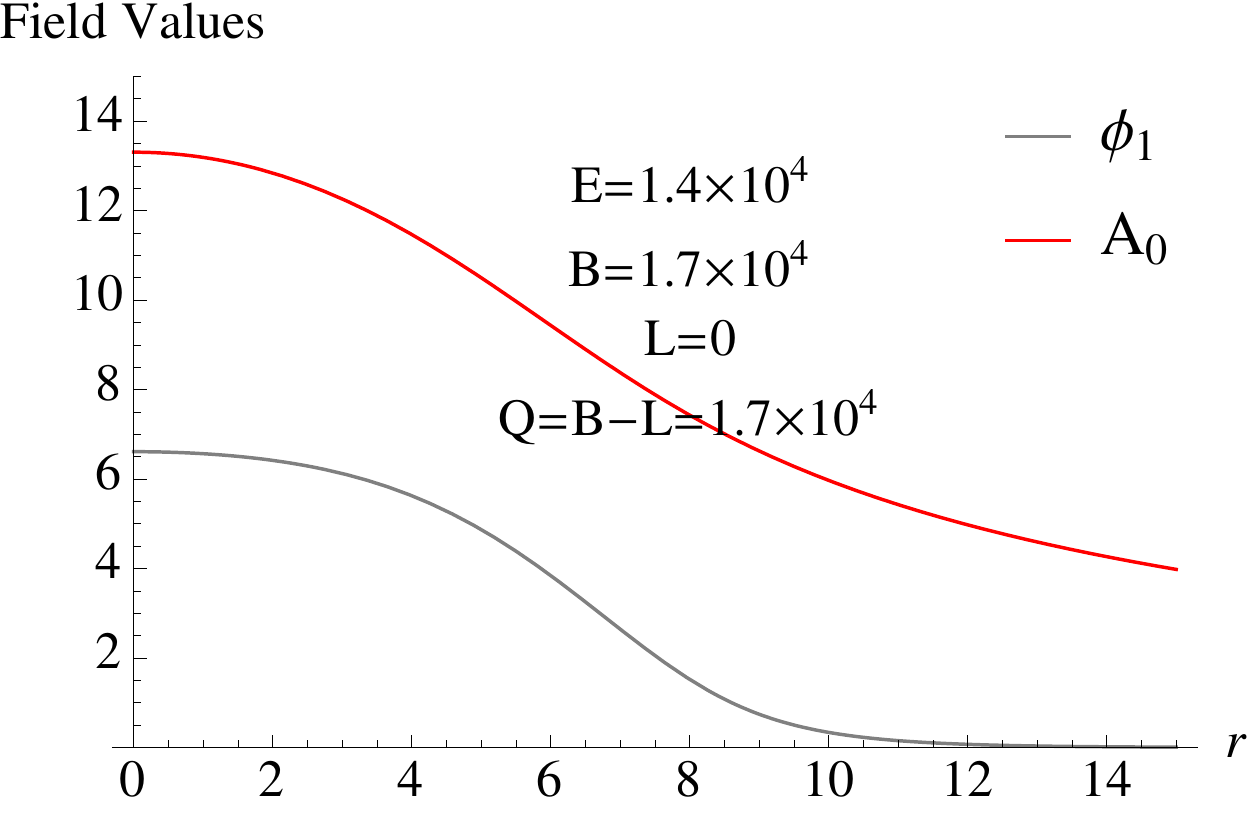}
 \end{minipage}
 \caption{2-scalar gauged Q-balls in gauge mediation model with $e^2=0.002$, $m_\Phi=1$ and $B=1.7\times10^4$.}
\label{fig:pptf}
\end{figure}

\begin{figure}[!h]
\begin{minipage}{.45\linewidth}
  \includegraphics[width=\linewidth]{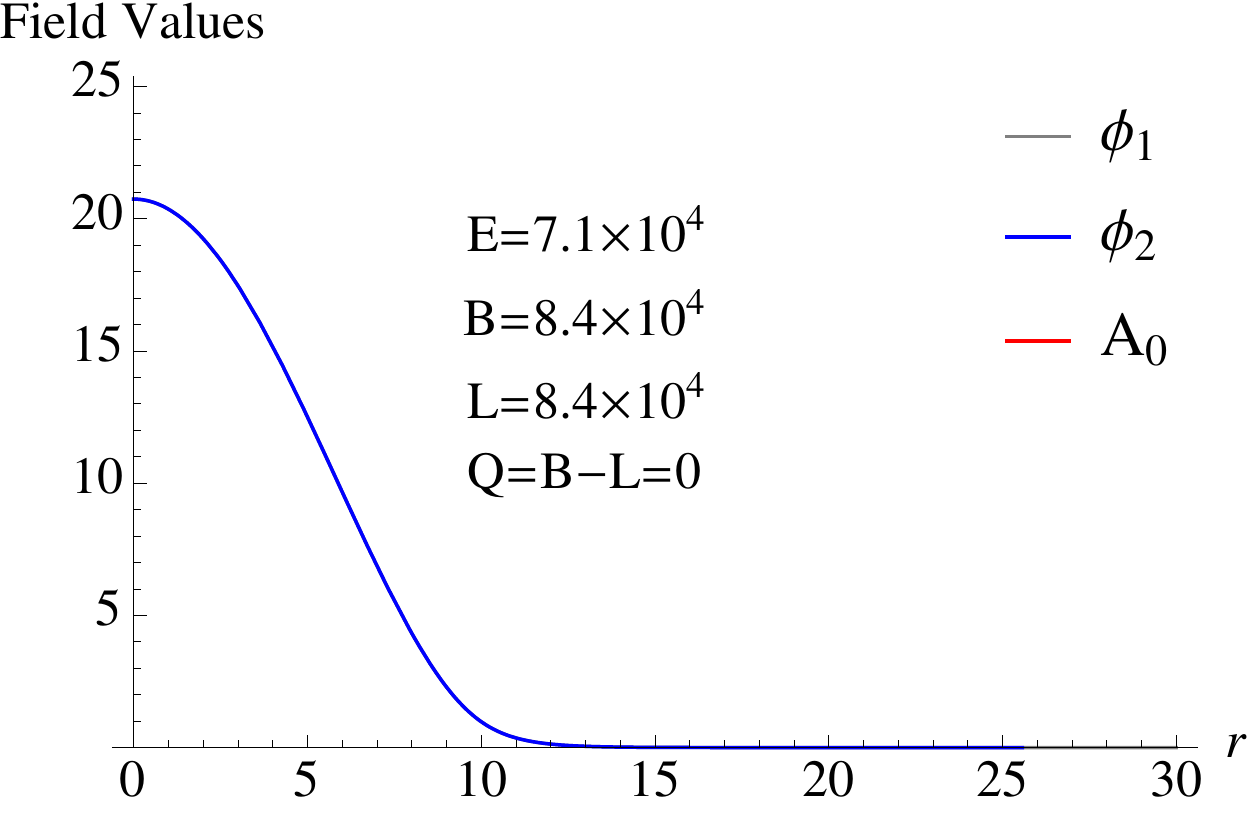}
 \end{minipage}
 \hspace{1cm}
 \begin{minipage}{.45\linewidth}
  \includegraphics[width=\linewidth]{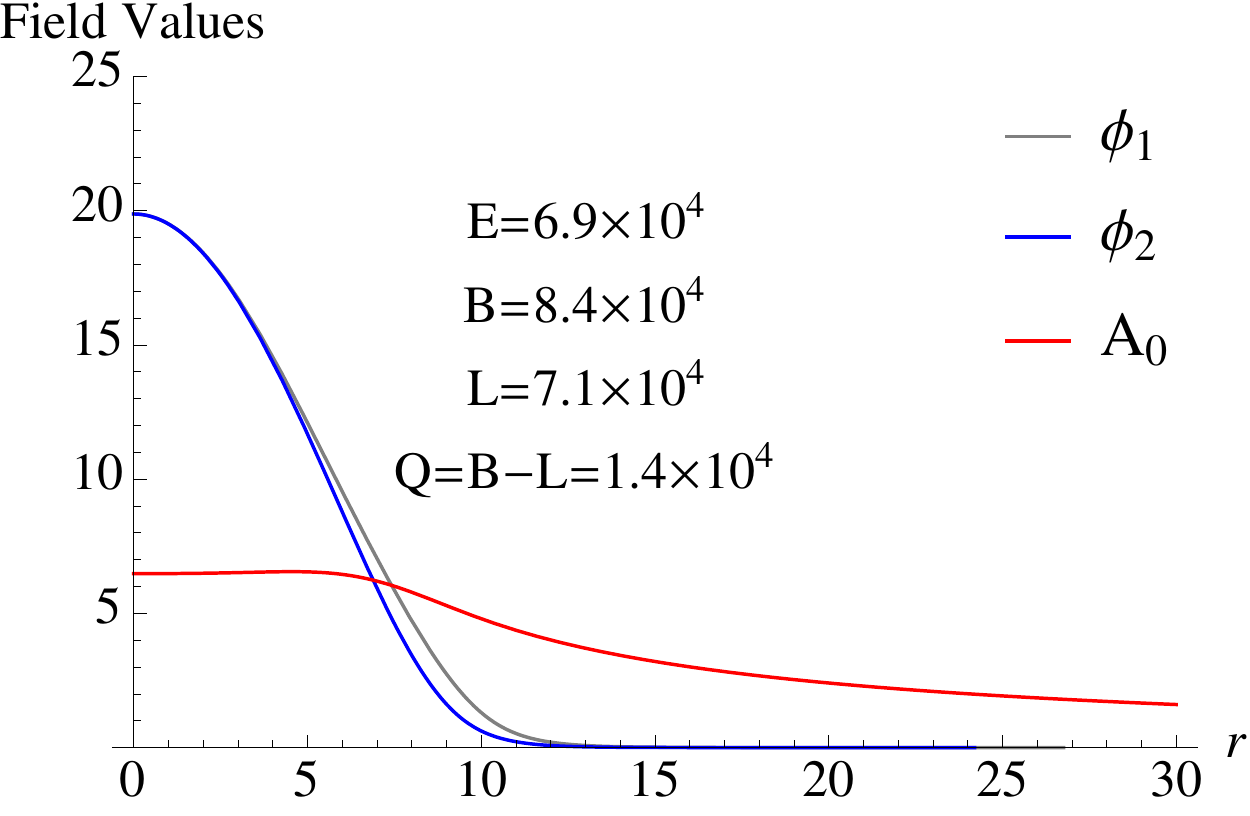}
 \end{minipage}
 \begin{minipage}{.45\linewidth}
  \includegraphics[width=\linewidth]{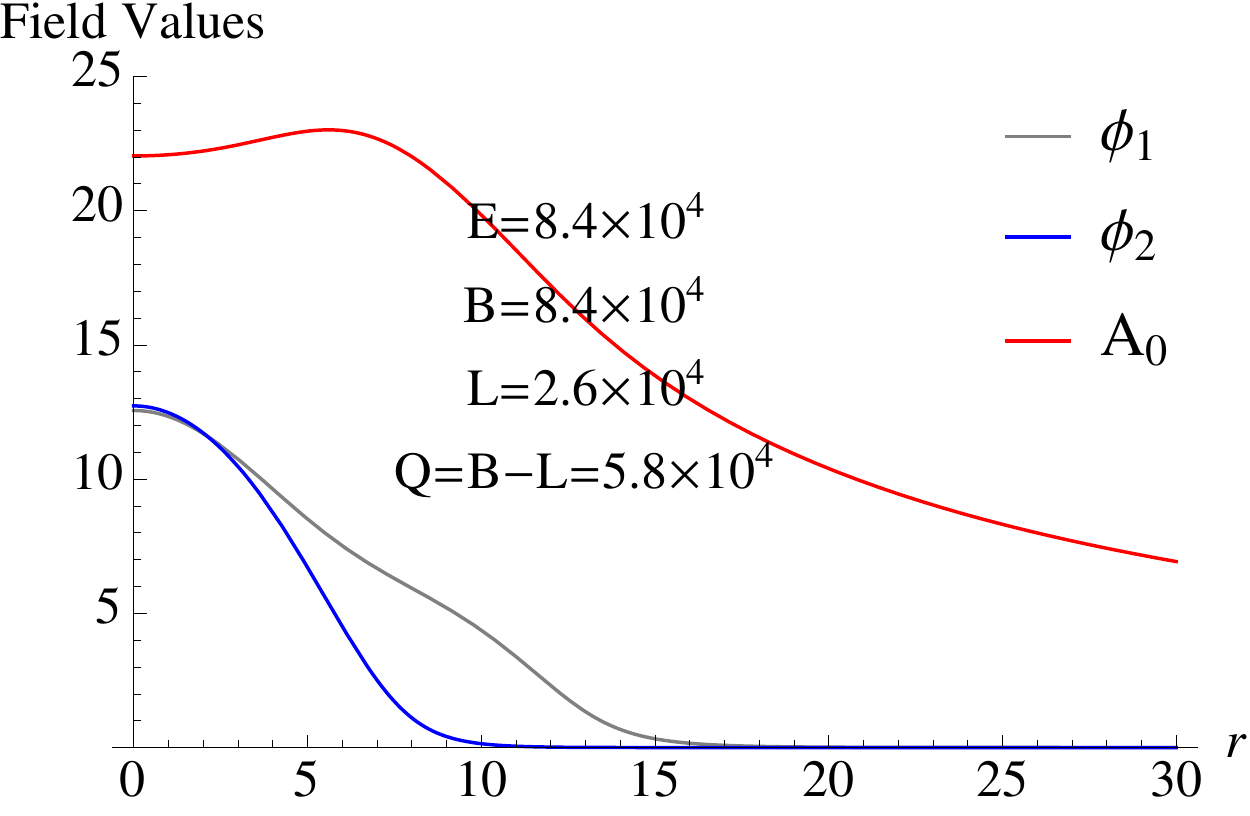}
 \end{minipage}
 \hspace{1cm}
 \begin{minipage}{.45\linewidth}
  \includegraphics[width=\linewidth]{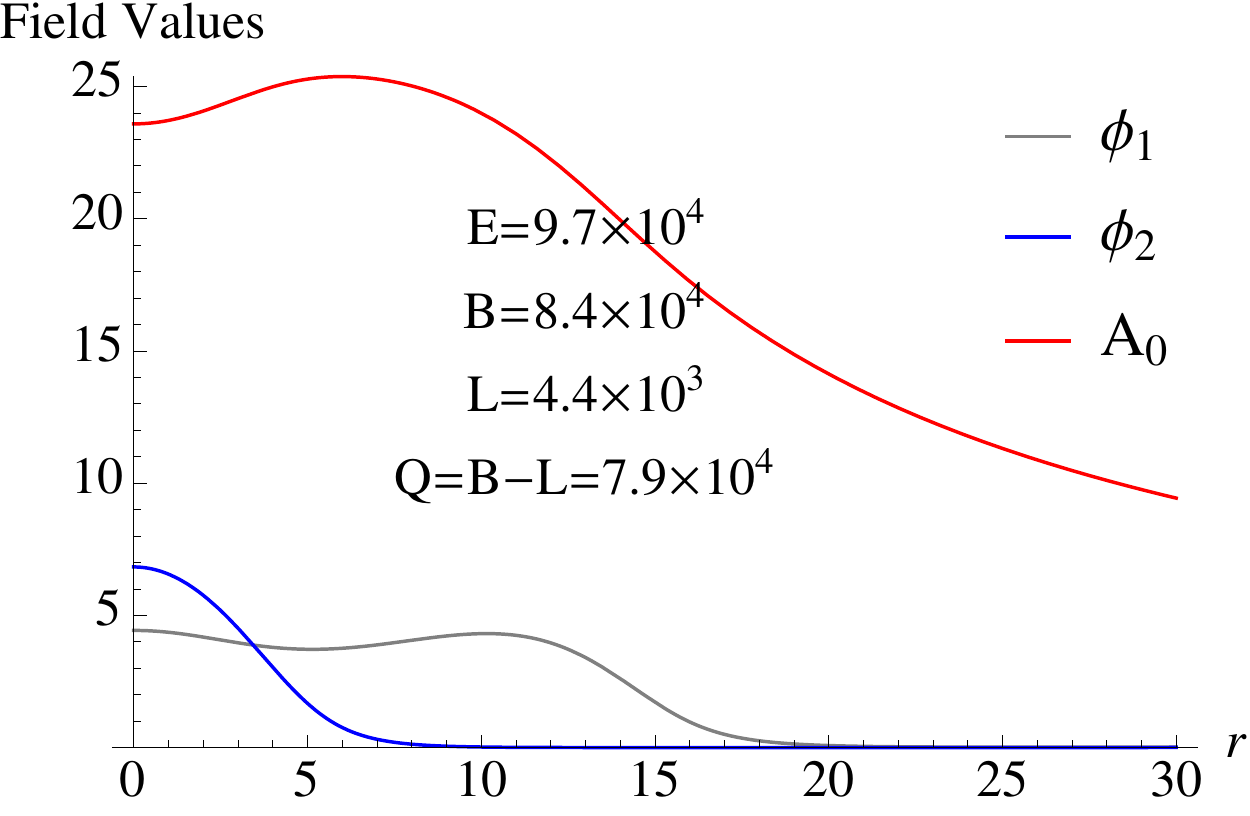}
 \end{minipage}
 \begin{minipage}{.45\linewidth}
  \includegraphics[width=\linewidth]{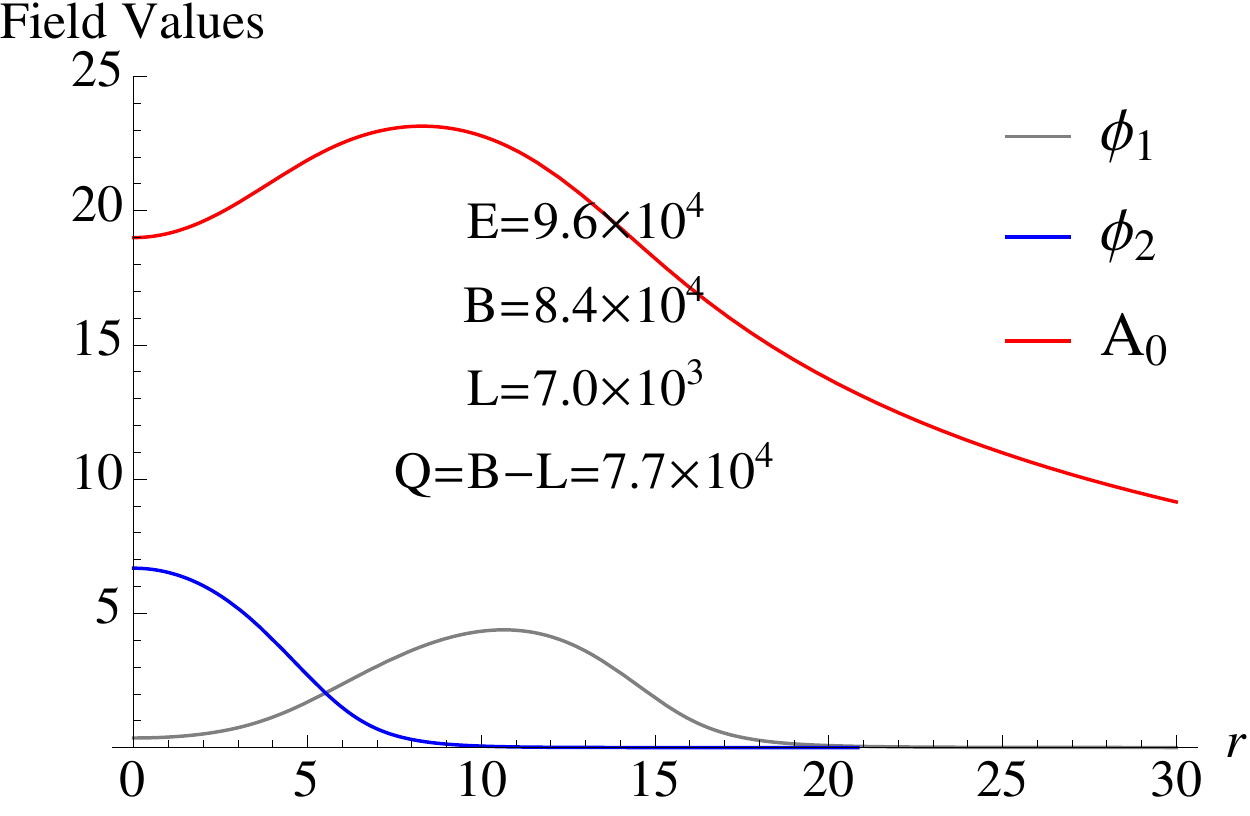}
 \end{minipage}
 \hspace{1.4cm}
 \begin{minipage}{.45\linewidth}
  \includegraphics[width=\linewidth]{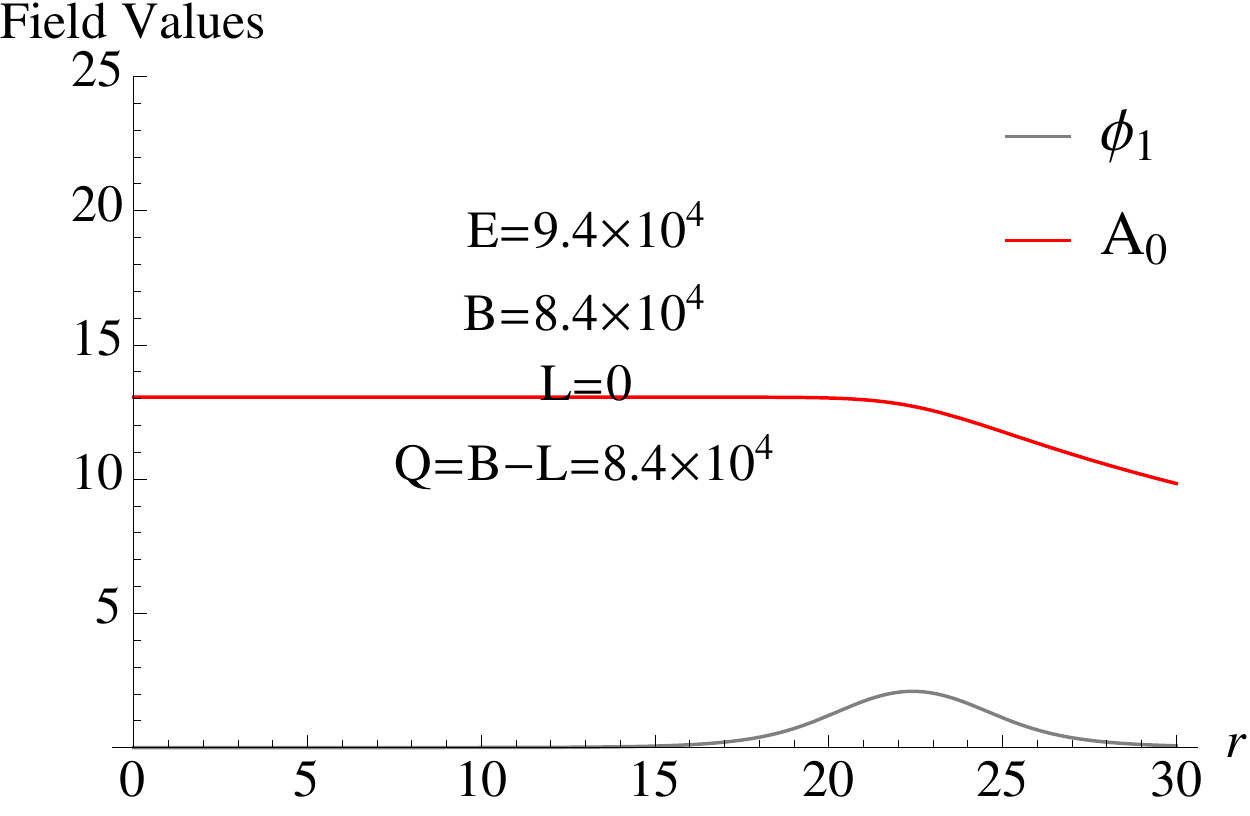}
 \end{minipage}
 \caption{2-scalar gauged Q-balls in gauge mediation model with $e^2=0.002$, $m_\Phi=1$ and $B=8.4\times10^4$.}
\label{fig:pptf2}
\end{figure}

We assume that the decay of the Q-ball always takes place if energetically allowed and that the evolution of the Q-ball can be approximated as a sequence of gauged Q-ball solutions. Then the decay of the leptonic component only can be represented by a sequence of gauged Q-ball solutions with the same baryon number arranged in descending order of lepton number, which is expected to decrease due to the decay. The results for $B=1.7\times10^4$ and $B=8.4\times10^4$ are shown in Fig.~\ref{fig:pptf} and in Fig.~\ref{fig:pptf2} respectively, where we also used the following approximate form of the gauge mediation potential: 
\begin{align}
V(\Phi_1,\Phi_2)=m_{\Phi_1}^4\ln\left(1+\frac{|\Phi_1|^2}{m_{\Phi_1}^2}\right)+m_{\Phi_2}^4\ln\left(1+\frac{|\Phi_2|^2}{m_{\Phi_2}^2}\right)+\frac{e^2}2\left(|\Phi_1|^2-|\Phi_2|^2\right)^2.
\end{align}
Here, we include the D-term potential, which arises since the D-flat condition $|\Phi_1|=|\Phi_2|$ is not valid anymore, and in addition, we assume $m_{\Phi_1}=m_{\Phi_2}\equiv m_{\Phi}$. We see that as the leptonic component decays, the gauge field, or the Coulomb potential arises. Therefore, it is expected that through decay process, the initially formed neutral Q-ball may evolve into the charged or gauged Q-ball, which means that the charged Q-balls may emerge in our universe. Next, if we look at the energy, we see that in Fig.~\ref{fig:pptf}, the energy decreases along the decay, which indicates that the particle which comes out from the Q-ball has positive energy. This means that the Q-ball emits free (i.e. unbounded to the Q-ball) particles until the leptonic component completely vanishes. Whereas in Fig.~\ref{fig:pptf2}, the energy starts to increase in the middle of the decay, which means that the energy of the emitted particle becomes negative. This in turn means that the particle starts to be bound to the Q-ball. This may be understood in the context of quantum mechanics of many-body system as a cloudy bound state which is analogous to that of an atomic system. We also see a slight decrease in energy near the end of the decay, which means that the emitted particle becomes free again. This is interpreted as follows. As shown in Fig.~\ref{fig:pptf2}, the leptonic component concentrates at the center while the baryonic one locates outside away from the center. Since the baryonic component is far enough from the surface of the leptonic component, from which the particle is emitted, the particle is initially accelerated outward enough to escape from the Q-ball eventually. 
\begin{figure}[!h]
  \includegraphics[width=\linewidth]{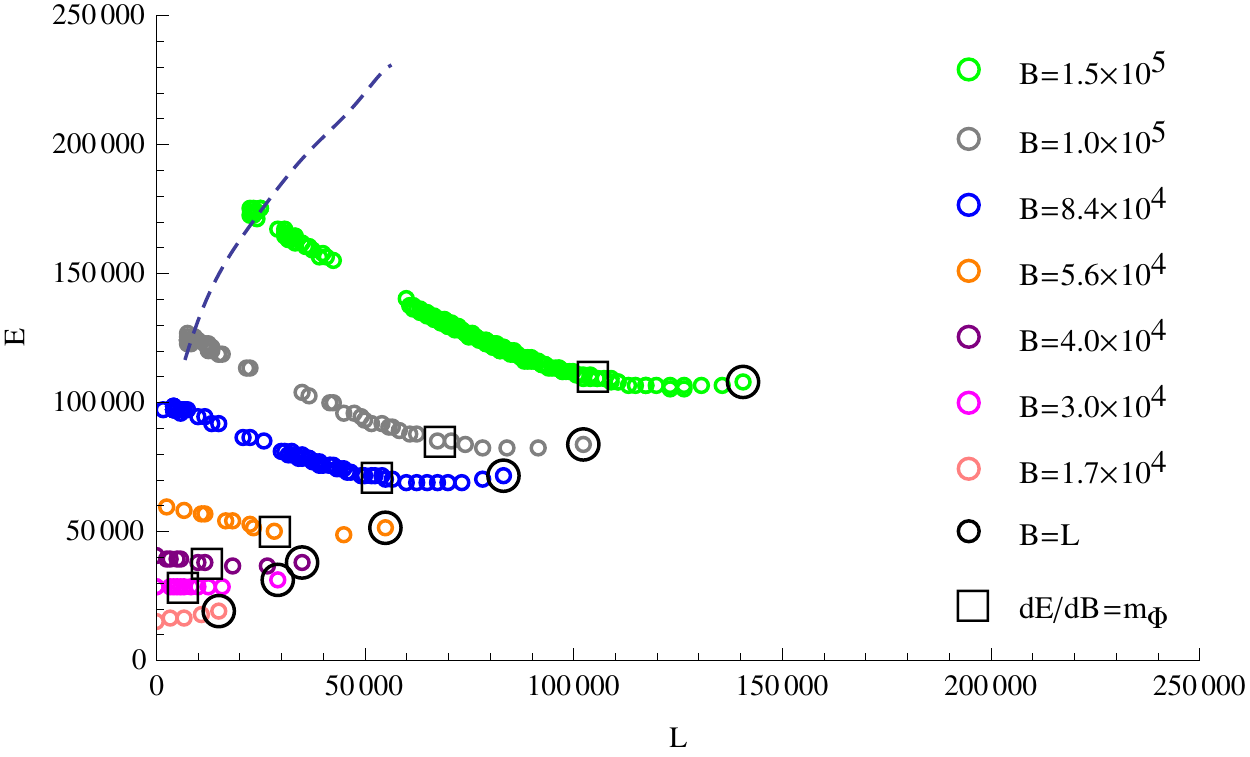}
 \caption{Solutions with various baryon numbers in the $E-L$ plane. The black circles are the solutions with $B=L$ and the black squares are the solutions with $(\del E/\del B)_L=m_\Phi$.}
\label{fig:kekka}
\end{figure}
\begin{figure}[t]
\begin{center}
  \includegraphics[width=0.7\linewidth]{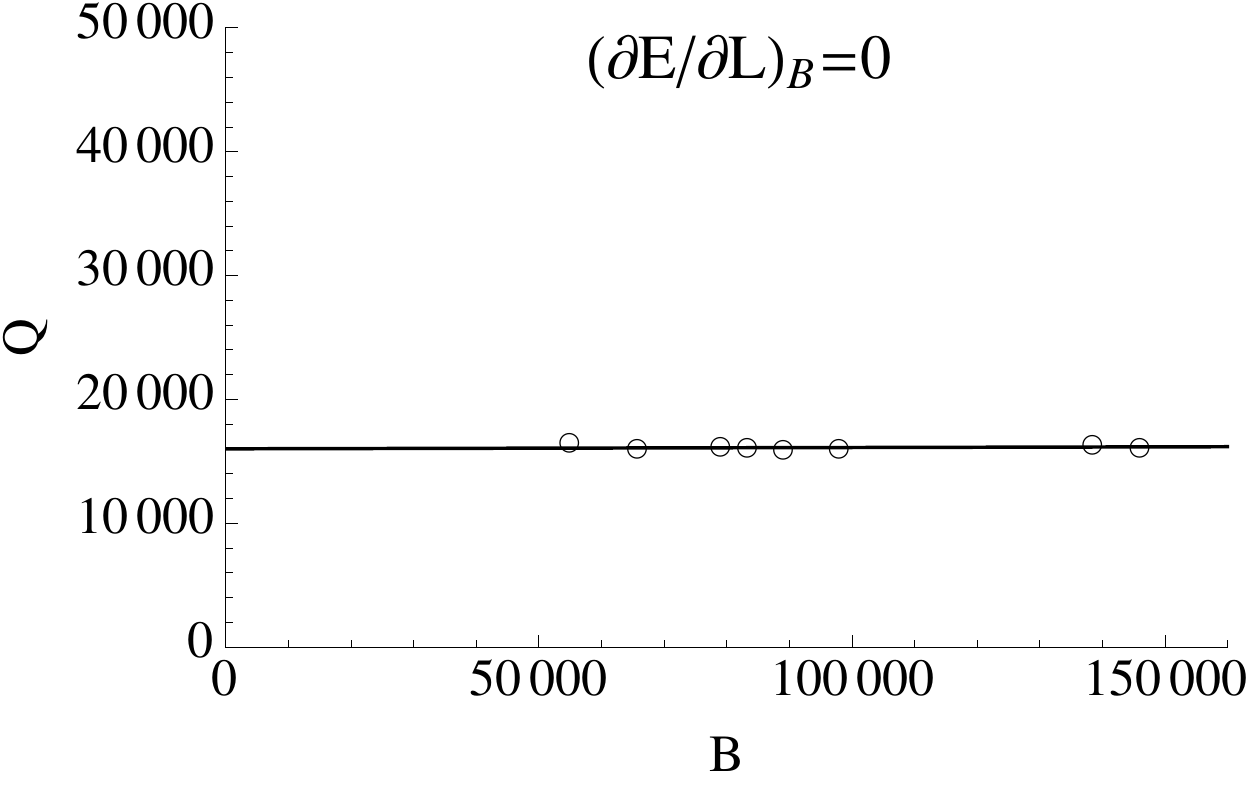}
 \caption{Electric charge $Q$ when $(\del E/\del L)_B=0$.}
\label{fig:2sdedl}
\end{center}
\end{figure}
\begin{figure}[t]
\begin{center}
  \includegraphics[width=0.7\linewidth]{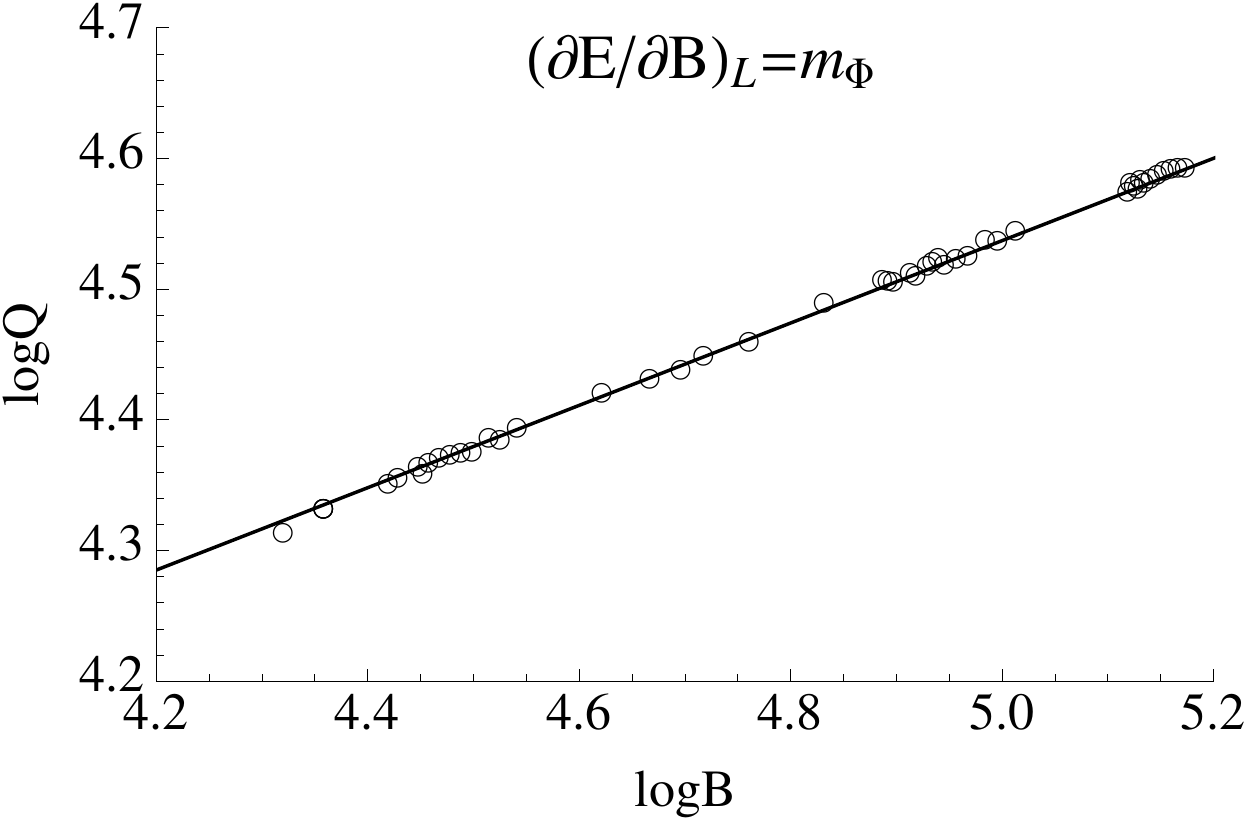}
 \caption{Electric charge when $(\del E/\del B)_L=m_\Phi$.}
\label{fig:2sdedb}
\end{center}
\end{figure}

We show the solutions with various baryon numbers in Fig.~\ref{fig:kekka}.\footnote{The gaps in some data are due to the switching of algorithm for the computation.} The black circles indicates the solutions with $B=L$, which are realized at the Q-ball formation after the Affleck-Dine mechanism\footnote{In the previous studies, the solutions were assumed to remain to be $B=L$, except for Ref.~\cite{ktt}.}. From there, the energy decreases along with $L$, which means that free particles are emitted and the Q-ball becomes electrically charged. The emission continues until the leptonic component vanishes for $B\lesssim3\times10^4$. However, for $B\gtrsim3\times10^4$, the energy starts to increase in the middle of the decay, which means that the particle starts to form a cloud of bounded leptonic particles. In Fig.~\ref{fig:2sdedl}, we show the electric charge of Q-ball at which $(\del E/\del L)_B=0$. From the figure, $(\del E/\del L)_B=0$ occurs for 
\begin{align}
Q\simeq1.6\times10^4
\end{align} 
and it is independent of baryon charge. This can be understood in the following way. When we approximate the energy of an emitted leptonic particle $(\del E/\del L)_B$ as the energy with electricity switched off plus Coulomb energy, 
\begin{align}
\left(\frac{\del E}{\del L}\right)_B&\approx\omega_{0L}-\frac{e^2Q}{4\pi R_L}\nonumber\\
&\simeq\omega_{0L}-\frac{e^2Q}{4\pi^2}\omega_{0L}
\label{eq:llo}
\end{align}
where we used the analytic expression with $R=\pi/\omega_{0}$. Equating $(\del E/\del L)_B$ to zero, we obtain
\begin{align}
Q=\frac{4\pi^2}{e^2}\simeq2.0\times10^4.
\end{align}
This roughly explains Fig.~\ref{fig:2sdedl}. 

As in the case of 1-scalar gauged Q-ball, it is possible that 
\begin{align}
\left(\partial{E}/\partial B\right)_L> m_\Phi
\end{align}
due to the electrical repulsion. In order to check this, we only need to examine $\omega_1$ of each solution since $(\del E/\del B)_L=\omega_1$ as proven above. We illustrate where $\left(\partial{E}/\partial B\right)_L=m_\Phi$ by black squares in Fig.~\ref{fig:kekka}. We plot the electric charge $Q$ at which $(\del E/\del B)_L$ becomes $m_\Phi$ for each $B$ in Fig.~\ref{fig:2sdedb}. The results can be fitted as 
\begin{align}
\log_{10} Q=3.0+3.0\times10^{-1}\log_{10} B
\end{align} 
This can be explained as follows. In the same way as deriving Eq.~(\ref{eq:llo}), the condition $(\del E/\del B)_L=m_\Phi$ is written as
\begin{align}
m_\Phi&=\left(\frac{\del E}{\del B}\right)_L\label{eq:dedbph}\\
&\simeq \omega_{0B}+\frac{e^2Q}{4\pi R_B}\nonumber\\
&=\omega_{0B}+\frac{e^2Q}{4\pi^2}\omega_{0B}\nonumber\\
&\simeq\frac{e^2Q}{4\pi^2}\omega_{0B}\nonumber\\
&=\frac{e^2Q}{2\sqrt{2}\pi}V_0^{1/4}B^{-1/4}.
\end{align}
Thus, we obtain
\begin{align}
Q=\frac{2\sqrt{2}\pi m_\Phi}{V_0^{1/4}e^2}B^{1/4},
\end{align}
which leads to
\begin{align}
\log_{10} Q&=\log_{10}\left(\frac{2\sqrt{2}\pi m_\Phi}{V_0^{1/4}e^2}\right)+\frac14\log_{10} B\nonumber\\
&\simeq3.4+2.5\times10^{-1}\log_{10} B,
\end{align}
where we used the same approximation as before. This roughly explains our numerical results.

The decay into protons is also expected to occur due to the electrical repulsion, even if we assume that the initial neutral Q-ball is stable against it. If this happens, baryonic component also decays off, and therefore charged Q-balls cannot be left in the universe. However, if the leptonic decay stops before the electric charge becomes large enough for the baryonic decay to occur, the evolution stops and charged Q-balls may survive. One way to stop the leptonic decay is that the leptonic cloud is close enough to the surface of the Q-ball so that the particle cannot come out. We derive the condition that this happens before the baryonic decay occurs. Let us suppose that the emitted leptons are electrons and roughly estimate the size of the leptonic cloud as Bohr radius. Then the condition that the cloud radius becomes equal to the size of the Q-ball is written as\begin{align}
\frac{4\pi}{m_eQe^2}&=R=\frac{\pi}{\omega_0}=\frac1{\sqrt{2}} V_0^{-1/4}B^{1/4},\\
Q&=\frac{4\sqrt{2}\pi V_0^{1/4}B^{-1/4}}{m_ee^2}.
\end{align}
This must happen for 
\begin{align}
Q<\frac{2\sqrt{2}\pi m_p}{V_0^{1/4}e^2}B^{1/4},
\end{align}
that is, before the baryonic decay starts to occur, where we have used the analysis parallel to that below Eq.~(\ref{eq:dedbph}). Therefore, the condition that the leptonic decay stops before baryonic decay occurs is given by
\begin{align}
B&>\frac{4V_0}{m_e^2m_p^2}\sim10^{30}\left(\frac{V_0}{(10^6\mathrm{GeV})^4}\right).
\end{align}
This implies that if the baryon number is large enough, the leptonic decay may stop before the baryonic decay takes place, so that the charged Q-ball may survive as a relic in the universe.

We see that for 
\begin{align}
B>\frac{4V_0}{\pi^4m_e^4}\sim10^{35}\left(\frac{V_0}{(10^6\mathrm{GeV})^4}\right),
\end{align}
Bohr radius is already smaller than the Q-ball size when the cloud is about to be formed, which means that the evolution stops without forming the cloud. Thus, more accurately, if 
\begin{align}
\frac{4V_0}{m_e^2m_p^2}<B<\frac{4V_0}{\pi^4m_e^4},
\end{align}
then the charged Q-balls are expected to survive with a cloud surrounding it, and if 
\begin{align}
B>\frac{4V_0}{\pi^4m_e^4},
\end{align}
then the charged Q-balls are expected to survive without the cloud\footnote{If $B>4V_0/\pi^4m_e^4$, the baryonic decay starts to occur when $Q=4m_p/ m_ee^2$, and this is larger than $Q=4\pi^2/e^2$, which is when the evolution is expected to stop. Thus, there is no need to worry that the baryonic decay occurs before the evolution stops.}.

Finally, the dashed line in Fig.~\ref{fig:kekka} illustrates $Q_{max}$ for each $B$, above which Q-ball solutions cannot exist. This results from electrical repulsion due to the gauge field.
\section{Conclusions and discussion}
\label{sec:conc}
In this paper, we considered gauged Q-balls in the two scalar model in order to discuss the evolution of neutral Q-balls which are formed from the flat direction in the Affleck-Dine mechanism and the possibility of realization of gauged Q-balls during their evolution. We approximated the evolution as a sequence of charged Q-ball solutions, which is implied from the situation that only the leptonic component decays off. As a result, a Coulomb potential arises and the Q-ball becomes electrically charged as expected. In other words, it is energetically favored for leptonic decay to occur. However, since there is an upper bound on charge of gauged Q-ball, the amount of decay is limited as well, which we examined quantitatively. In addition, if the baryon number of the initially formed Q-ball is large enough, the electric charge of the Q-ball grows enough so that the particle which is emitted from the Q-ball is bound to it. The baryonic decay is also expected to occur by virtue of the electrical repulsion, which leads to the vanishing of the charged Q-balls in the universe. However, it is expected that if the leptonic cloud is close enough to the surface of the Q-ball, the leptonic decay can stop before the electric charge becomes large enough for the baryonic decay to occur, so that the evolution stops and charged Q-balls can survive. We roughly estimated when this can happen, and as a consequence, we found that there exists a lower bound on the baryon number. 

Suppose that dark matter consists of Q-balls. Since a Q-ball is known to absorb protons and emit pions, the neutral Q-balls can be detected by Super-Kamiokande~\cite{sipal, kensyutu, kttn} or IceCube~\cite{icecube}, while these detectors are not suited for detection of the charged Q-balls since charged Q-balls cannot absorb the protons due to the electrical repulsion. Thus, the electric charge is expected to make differences in the experimental signatures of the relic Q-balls. The charged Q-balls are expected to behave as some kind of nuclei, and are known to be detectable by such detectors as MACRO~\cite{sipal,macr} and the observational bounds on mass and flux of the relic charged Q-balls are obtained~\cite{kensyutu}. The emitted particles are also expected to contribute to the energy components of the universe and leptonic particles must satisfy the observational bounds on neutrino component from WMAP~\cite{neutrinof}, which must be verified in future work.

Although we assumed that the decay occurs if energetically allowed, we did not specify which kind of decay we are considering, for instance species of particle which is emitted, an actual decay rate etc. These informations must be added when we consider electrodynamical effects and actually solve time development of Q-ball. Lastly, here we are considering the simplest model of a flat direction, but in reality, we may need to consider more complex flat directions, possibly in nonabelian gauge theories as well.

\vspace{1cm}

\section*{Acknowledgments}
This work is supported by Grant-in-Aid for Scientific research 
from the Ministry of Education, Science, Sports, and Culture
(MEXT), Japan, No. 25400248 (M.K.), 
World Premier International Research Center Initiative
(WPI Initiative), MEXT, Japan,
and the Program for the Leading Graduate Schools, MEXT, Japan (M.Y.).
M.Y. acknowledges the support by JSPS Research Fellowships for Young Scientists, No.25.8715.

\vspace{1cm}



\end{document}